\documentclass[12pt,onecolumn]{IEEEtran}%

\usepackage{amsmath}
\usepackage{graphicx}
\usepackage{amscd}
\usepackage{amsfonts}
\usepackage{amssymb}%

\renewcommand{\baselinestretch}{1.2}  

\newtheorem{theorem}{Theorem}

\newtheorem{corollary}[theorem]{Corollary}

\newtheorem{definition}[theorem]{Definition}
\newtheorem{example}[theorem]{Example}

\newtheorem{lemma}[theorem]{Lemma}

\newtheorem{proposition}[theorem]{Proposition}

\newcommand{\var}{{\rm{Var}}}

\newcommand{\po}{{\rm{Po}}}
\newcommand{\Po}{\po}

\newcommand{\Cpo}{{\rm{CPo}}}

\newcommand{\blah}[1]{}

\def\be{\begin{eqnarray}}
\def\ee{\end{eqnarray}}
\def\ben{\begin{eqnarray*}}
\def\een{\end{eqnarray*}}

\begin{document}

\title{Thinning, Entropy and the Law of Thin Numbers}
\author{Peter Harremo\"{e}s\thanks{P.
Harremo\"{e}s was supported by a grant from the Danish Natural Science
Research Council and the European Pascal Network of Excellence.},
Oliver Johnson and Ioannis Kontoyiannis
\thanks{I. Kontoyiannis was supported in part by 
a Marie Curie International Outgoing Fellowship.}
\thanks{P.
Harremo\"{e}s is with Inst. Mathematics and Comp. Science, Amsterdam, 1090
GB, The Netherlands, P.Harremoes@cwi.nl. Oliver Johnson is with Dept.
Mathematics, University of Bristol, Bristol, BS8 1TW, United Kingdom,
O.Johnson@bristol.ac.uk. Ioannis Kontoyiannis is with the Department of
Informatics, Athens University of
Economics \& Business, Patission 76, Athens 10434, Greece, yiannis@aueb.gr.}}
\maketitle

\begin{abstract}
R\'enyi's {\em thinning} operation on a discrete random variable 
is a natural discrete analog of the scaling operation for 
continuous random variables. The properties of thinning 
are investigated in an information-theoretic context,
especially in connection with information-theoretic inequalities
related to Poisson approximation results. The classical 
Binomial-to-Poisson convergence (sometimes referred to 
as the \textquotedblleft law of small numbers\textquotedblright) 
is seen to be a special case of a thinning limit theorem 
for convolutions of discrete distributions. A rate of convergence 
is provided for this limit, and nonasymptotic bounds are also
established. This development parallels, in part, the 
development of Gaussian inequalities leading to the 
information-theoretic version of the central limit theorem.
In particular, a ``thinning Markov chain'' is introduced,
and it is shown to play a role analogous to that 
of the Ornstein-Uhlenbeck process in connection to the
entropy power inequality.
\end{abstract}


\newpage

\section{Introduction}
\label{sec:intro}

Approximating the distribution of a sum of weakly 
dependent discrete random
variables by a Poisson distribution is an important
and well-studied problem in probability;
see \cite{BarHolJan92} and the references therein
for an extensive account. Strong connections between
these results and information-theoretic techniques 
were established \cite{Harremoes01}\cite{KonHarJoh03}. 
In particular, for the special case of approximating 
a binomial distribution by a Poisson, some of the sharpest 
results to date are established using a combination 
of the techniques \cite{Harremoes01}\cite{KonHarJoh03} 
and Pinsker's inequality \cite{Csiszar67}%
\cite{Fedotovetal03}\cite{HarRuz}.
Earlier work on information-theoretic bounds 
for Poisson approximation is reported in 
\cite{Zolotarev1983}\cite{johnstone-macginbbon:87}\cite{Reiss1989}.

The thinning operation, which we define next, was introduced by 
R\'{e}nyi in \cite{Renyi1956}, who used it to provide an 
alternative characterization of Poisson measures.

\medskip

\begin{definition}
\label{def:thin} Given $\alpha\in\lbrack0,1]$ and a discrete random variable
$X$ with distribution $P$ on $\mathbb{N}_{0}=\{0,1,\ldots\}$, the
{\em $\alpha$-thinning of $P$} is the distribution $T_{\alpha}(P)$ of the sum,
\begin{equation}
\sum_{x=1}^{X}B_{x},\;\;\;\;\text{where}\;B_{1},B_{2}\ldots\sim\mbox{i.i.d.\
Bern$(\alpha)$},\label{eq:intro}%
\end{equation}
where the random variables $\{B_x\}$ are independent and identically
distributed (i.i.d.) each with a Bernoulli distribution with parameter
$\alpha$, denoted Bern($\alpha$), and also independent of $X$.
[As usual, we take the empty sum $\sum_{x=1}^0(\cdot)$ to be
equal to zero.]
An explicit representation of $T_{\alpha}(P)$ can be given as,%
\begin{equation}
T_{\alpha}(P)  (  z)  =
\sum_{x=z}^{\infty}P(x)  
\binom{x}{z}\alpha^{z}(  1-\alpha)  ^{x-z},
\;\;\;\;z\geq
0.\label{eq:intro2}%
\end{equation}
When it causes no ambiguity, the thinned distribution
$T_\alpha(P)$ is written simply $T_\alpha P$.
\end{definition}

\medskip

For any random variable $X$ with distribution $P$ on $\mathbb{N}_{0}$, we
write $P^{\ast n}$ for the $n$-fold convolution of $P$ with itself, i.e., the
distribution of the sum of $n$ i.i.d.\ copies of $X$. 
For example, if $P\sim$ Bern$(p)$, 
then $P^{\ast n}\sim\mathrm{{Bin}}(n,p)$,
the binomial distribution with parameters $n$ and $p$.
It is easy to see that its $(1/n)$-thinning,
$T_{1/n}(P^{\ast n})$, is simply
$\mathrm{{Bin}}(n,p/n)$; 
see Example~\ref{ex:BernS} below. Therefore, the classical 
Binomial-to-Poisson convergence result -- sometimes
referred to as the ``law of small numbers'' --
can be phrased as saying that, if $P\sim$ Bern($p$), then,
\begin{eqnarray}
T_{1/n}(P^{\ast n})\rightarrow\Po(p),\;\;\;\text{as}\;n\rightarrow
\infty,
\label{eq:lsn}
\end{eqnarray}
where Po($\lambda$) denotes the Poisson
distribution with parameter $\lambda>0$.

One of the main points of this work is to show 
that this result holds for very wide class 
of distributions $P$, and to provide conditions
under which several stronger and more general versions 
of (\ref{eq:lsn}) can be obtained. We refer to results 
of the form (\ref{eq:lsn}) as {\em laws of thin numbers}.

Section~\ref{sec:exclass} contains numerous examples that
illustrate how particular families of random 
variables behave on thinning, and it also introduces
some of the particular classes of random variables that
will be considered in the rest of the paper.
In Sections~\ref{sec:ltn} 
and~\ref{sec:ltn2} several versions of the 
law of thin numbers are formulated;
first for i.i.d.\
random variables in Section~\ref{sec:ltn},
and then for general classes of (not necessarily
independent or identically distributed)
random variables in Section~\ref{sec:ltn2}. 
For example, in the simplest case where
$Y_1,Y_2,\ldots$ are i.i.d.\
with distribution $P$ on $\mathbb{N}_{0}$ and 
with mean $\lambda$, so that 
the distribution of their sum,
$S_n=Y_1+Y_2+\cdots+Y_n$, is $P^{*n}$,
Theorem~\ref{thm:ltnstrong}
shows that,
\begin{eqnarray}
D\left(  T_{1/n}(  P^{\ast n})
\Vert\Po(  \lambda)  \right)  \to 0,
\;\;\;\;\mbox{as}\;n\to\infty,
\label{eq:LTN1}
\end{eqnarray}
as long as $D(  P\Vert\Po(  \lambda)  )  <\infty$,
where, as usual, $D(P\Vert Q)$ 
denotes the {\em information divergence},
or {\em relative entropy}, from $P$ to $Q$,\footnote{Throughout 
the paper, $\log$ denotes
the natural logarithm to base $e$, and we adopt the usual
convention that $0\log 0 = 0$.}
\[
D(  P\Vert Q)  =\sum_{z=0}^{\infty}P(z)  \log
\frac{P(z)}{Q(z)}.
\]
Note that, unlike most classical Poisson
convergence results, the
law of thin numbers in (\ref{eq:LTN1})
proves a Poisson limit theorem
for the sum of a single sequence
of random variables, rather than for a triangular array. 

It may be illuminating to compare the result (\ref{eq:LTN1})
with the information-theoretic version of the central
limit theorem (CLT); see, 
e.g., \cite{Barron86}\cite{Johnson04}.
Suppose 
$Y_1,Y_2,\ldots$ are i.i.d.\
continuous random variables with density
$f$ on $\mathbb R$, and with zero mean
and unit variance. Then the density of their sum
$S_n=Y_1+Y_2+\cdots+Y_n$,
is the $n$-fold convolution 
$f^{*n}$ of $f$ with itself.
Write $\Sigma_\alpha$ for the standard scaling 
operation in the CLT regime: If a continuous random 
variable $X$ has density $f$, then $\Sigma_\alpha(f)$ 
is the density of the scaled random variable $\sqrt{\alpha} X$,
and, in particular, the density of the standardized
sum $\frac{1}{\sqrt{n}}S_n$ is $\Sigma_{1/n}(f^{*n})$.
The information-theoretic CLT
states that, if $D(f\Vert\phi )  <\infty$,
we have,
\begin{eqnarray}
D\left(\left.  \Sigma_{1/n}(f^{\ast n})
\right\Vert\phi \right)  \to 0,
\;\;\;\;\mbox{as}\;n\to\infty,
\label{eq:CLT}
\end{eqnarray}
where $\phi$ is the standard Normal 
density.
Note the close analogy between the 
statements of the
law of thin numbers 
in~(\ref{eq:LTN1}) and the CLT
in~(\ref{eq:CLT}).

Before describing the rest of our results,
we mention that there is a significant thread
in the literature on thinning limit theorems
and associated results for point processes.
Convergence theorems of the ``law of thin
numbers'' type, as in (\ref{eq:lsn}) and (\ref{eq:LTN1}),
were first examined in the context of queueing theory
by Palm \cite{palm:43} 
and Khinchin \cite{khinchine:60},
while more general 
results were established by 
Grigelionis \cite{grigelionis:63}. 
See the discussion in the text,
\cite[pp.~146-166]{daley-vere-jones:II},
for details and historical remarks;
also see the comments following Theorem~\ref{thm:ltnweak2}
in Section~\ref{sec:ltn2}.
More specifically, this line of work considered
asymptotic results, primarily in the sense of
weak convergence, for the distribution of
a superposition of the sample paths of
independent (or appropriately weakly
dependent) point processes.
Here we take a different direction
and, instead of considering the full 
infinite-dimensional distribution 
of a point process,
we focus on finer results -- 
e.g., convergence in information
divergence and non-asymptotic bounds --
for the one-dimensional distribution of
the thinned sum of integer-valued 
random variables.

With these goals in mind,
before examining the finite-$n
$ behavior of $T_{1/n}(P^{*n})$, in Section~\ref{sec:mc} 
we study a simpler but related problem, 
on the convergence of a continuous-time
``thinning''
Markov chain on ${\mathbb N}_0$. 
In the present context,
this chain plays a role parallel to that 
of the Ornstein-Uhlenbeck process in the 
context of Gaussian convergence and the entropy 
power inequality
\cite{Shannon48}\cite{Stam59}\cite{Lieb78}. 
We show that the thinning Markov chain has the 
Poisson law as its unique invariant measure,
and we establish its convergence both in 
total variation and in terms
of information divergence. Moreover,
in Theorem~\ref{thm:chisquare}
we characterize precisely the rate
at which it converges to the Poisson law
in terms of the $\chi^2$ distance,
which also leads to an upper bound
on its convergence in information divergence.
A new characterization of the Poisson distribution 
in terms of thinning is also obtained. 
The main technical tool
used here is based on an examination of
the $L^2$ properties of the Poisson-Charlier 
polynomials in the thinning context.

In Section~\ref{sec:ub} we give both asymptotic
and finite-$n$ bounds on the rate of convergence
for the law of thin numbers. Specifically,
we employ the {\em scaled Fisher information}
functional introduced in 
\cite{KonHarJoh03} to give precise,
explicit bounds on the divergence,
$D(T_{1/n}(P^{*n})\|\Po(\lambda))$.
An example of the type of result we prove is the following:
Suppose $X$ is an ultra bounded 
(see Definition~\ref{def:classes} in Section~\ref{sec:exclass}) 
random variable, with distribution $P$,
mean $\lambda$, and finite variance $\sigma^2\neq\lambda$.
Then,
\[
\limsup_{n\rightarrow\infty}n^{2}D\left(  T_{1/n}(  P^{\ast n})  
\Vert\Po(  \lambda)  \right)  \leq2c^{2},
\]
for a nonzero constant $c$ we explicitly identify;
cf.\ Corollary~\ref{cor:ub}.

Similarly, in Section~\ref{sec:tv} we give
both finite-$n$ and asymptotic bounds on the
law of small numbers in terms of the total variation
distance,
$\|T_{1/n}(P^{*n})-\Po(\lambda)\|$,
between 
$T_{1/n}(P^{*n})$ and the $\Po(\lambda)$
distribution. In particular, 
Theorem~\ref{2TV} states that if
$X\sim P$ has mean $\lambda$ and
finite variance $\sigma^2$, then,
for all $n$, 
\[
\left\Vert T_{1/n}(P^{\ast n})-\Po(\lambda)\right\Vert \leq\frac
{1}{n2^{1/2}}+\frac{\sigma}{n^{1/2}}\min\left\{  1,\frac{1}{2\lambda^{1/2}%
}\right\} .
\]

A closer examination of the monotonicity
properties
of the scaled Fisher information in relation
to the thinning operation is described
in Section~\ref{sec:monotone}.
Finally, 
Section~\ref{sec:comp} shows
how the idea 
of thinning can be
extended to compound Poisson 
distributions.
The Appendix contains the proofs of some of the more 
technical results.

Finally we mention that, after the announcement
of the present results in \cite{thinning:ISIT},
Yu \cite{yu:thinning:pre}
also obtained some interesting, related
results. In particular, he showed that the
conditions of the strong and
thermodynamic versions of the law
of thin numbers 
(see Theorems~\ref{thm:ltnstrong} 
and~\ref{thm:ltnthermo})
can be weakened,
and he also provided conditions
under which the convergence in these
limit theorems is monotonic in $n$.

\newpage

\section{Examples of Thinning and Distribution Classes\label{sec:exclass}}

This section contains several examples of the thinning operation, 
statements of its more basic properties, and the definitions
of some important classes of distributions that will be play 
a central role in the rest of this work. 
The proofs of all the
lemmas and propositions of this section are given in the Appendix.

Note, first, two important properties of thinning 
that are immediate from its definition: 

1.~The thinning of
a sum of independent random variables is the convolution 
of the corresponding thinnings. 

2.~For all $\alpha,\beta\in[0,1]$
and any distribution $P$ on ${\mathbb N}_0$, we have,
\begin{eqnarray}
T_\alpha(T_\beta(P))=T_{\alpha\beta}(P).
\label{eq:semigroup}
\end{eqnarray}

\begin{example}
Thinning preserves the Poisson law, in that 
$T_{\alpha}(\Po(\lambda))=\Po(\alpha\lambda)$.
This follows from~(\ref{eq:intro2}), since,
\begin{align*}
T_{\alpha}(  \Po(\lambda) )(z)   
&  =\sum_{x=z}^{\infty}\Po(  \lambda,x)
\binom{x}{z}\alpha^{z}(  1-\alpha)^{x-z}\\
&  =\sum_{x=z}^{\infty}\frac{\lambda^{x}}{x!}e^{-\lambda}\binom{x}{z}%
\alpha^{z}(  1-\alpha)^{x-z}\\
&  =\frac{e^{-\lambda}}{z!}(  \alpha\lambda)  ^{z}\sum
_{x=z}^{\infty}\frac{(\lambda(1-\alpha))^{x-z}%
}{(x-z)!}\\
&  =\frac{e^{-\lambda}}{z!}(  \alpha\lambda)  ^{z}e^{\lambda(
1-\alpha)  }\\
&  =\Po(\alpha\lambda,z),
\end{align*}
where
Po$(\lambda,x)=e^{-\lambda}\lambda^{x}/x!$, $x\geq 0$, denotes
the Poisson mass function.
\end{example}

\medskip

As it turns out, the factorial moments of 
a thinned distribution are easier to work with
than ordinary moments.
Recall that the {\em $k$th factorial moment}
of $X$ is 
$E[  X^{\underline{k}}]$,
where $x^{\underline{k}}$ denotes the falling 
factorial,
$$
x^{\underline{k}}\;=\;
x(x-1)  
\cdots(x-k+1)\;=\;\frac{x!}{(x-k)!}.$$

The factorial moments of an $\alpha$-thinning are easy to calculate:

\medskip

\begin{lemma}
\label{lem:scalmom} 
For any random variable $Y$ with distribution $P$ on
${\mathbb N}_0$ and for
$\alpha\in(0,1)$, writing $Y_{\alpha}$ for a random variable with distribution
$T_{\alpha}P$:
\begin{equation}
E[  Y_{\alpha}^{\underline{k}}]  =\alpha^{k}E[  Y^{\underline
{k}}]  .\;\;\;\mbox{ for all $k$.}
\end{equation}
That is, thinning scales factorial moments in the same way as ordinary
multiplication scales ordinary moments.
\end{lemma}

\medskip

We will use the following result, which is 
a multinomial version 
of Vandermonde's
identity and is easily proved by induction.
The details are omitted.

\medskip

\begin{lemma}
\label{lem:fallfact} The 
falling factorial satisfies the multinomial
expansion, i.e., for any positive integer $y$,
all integers $x_1,x_2,\ldots,x_y$, and any $k\geq 1$,
\[
\Big(\sum_{i=1}^{y}x_{i}\Big)^{\underline{k}}=\sum_{k_{1},k_2 \ldots,
k_{y}\;:\; k_{1} +k_2+ \ldots+ k_{y} =k}\binom{k}{%
\begin{array}
[c]{cccc}%
k_{1} & k_{2} & \cdots & k_{y}%
\end{array}
}%
{\displaystyle\prod\limits_{i=1}^{y}}
x_i^{\underline{k_i}}.
\]
\end{lemma}

\medskip

The following is a basic regularity property of the thinning 
operation.

\medskip

\begin{proposition}
\label{prop:inj}
For any $\alpha\in(0,1)$, the map $P\mapsto T_{\alpha}(P)$ is injective.
\end{proposition}

\medskip

\begin{example}
\label{ex:BernS}
Thinning preserves the class of Bernoulli sums. 
That is, the thinned version of
the distribution of a finite sum of independent Bernoulli random variables 
(with possibly different parameters) is also such a sum. This follows from 
property~1 stated in the beginning of this section, 
combined with the observation that the $\alpha$-thinning of the Bern($p$)
distribution is the Bern$(\alpha p)$ distribution.
In particular, thinning preserves the binomial family: 
$T_\alpha(\mbox{Bin}(n,p))=\mbox{Bin}(n,\alpha p)$.
\end{example}

\medskip

\begin{example}
Thinning by $\alpha$ transforms a geometric distribution with mean $\lambda$
into a geometric distribution with mean $\alpha\lambda$. Recalling that the
geometric distribution with mean $\lambda$ has point probabilities,%
\[
\mathrm{{Geo}}(\lambda,x)  =\frac{1}{1+\lambda}\Big(
\frac{\lambda}{1+\lambda}\Big)^x,\;\;\;\;\;
\;x = 0, 1, \ldots,
\]
using (\ref{eq:intro2}),
\begin{align*}
T_{\alpha}\mathrm{{Geo}}(\lambda)(z)   &
=\sum_{x=z}^{\infty}\frac{1}{1+\lambda}\left(  \frac{\lambda}{1+\lambda
}\right)  ^{x}\binom{x}{z}\alpha^{z}(1-\alpha)  ^{x-z}\\
&  =\frac{1}{(1+\lambda)  z!}\left(  \frac{\alpha\lambda
}{1+\lambda}\right)  ^{z}\sum_{x=z}^{\infty}\left(  \frac{\lambda(
1-\alpha)  }{1+\lambda}\right)  ^{x-z}x^{\underline{z}}\\
&  =\frac{1}{(  1+\lambda)  z!}\left(  \frac{\alpha\lambda
}{1+\lambda}\right)  ^{z}z!\left(  1-\frac{\lambda(1-\alpha)
}{1+\lambda}\right)  ^{-z-1}\\
&  =\mathrm{{Geo}}(\alpha\lambda,z).
\end{align*}
The sum of $n$ i.i.d.\ geometrics has
a negative binomial distribution. Thus, in view of this example
and property~1 stated in the beginning of this section,
the thinning of a negative binomial
distribution is also negative binomial.
\end{example}

\medskip

Partly motivated by these examples, we describe certain classes 
of random variables (some of which are new).
These appear as natural technical assumptions in
the subsequent development of our results. 
The reader may prefer to skip the remainder of 
this section and only refer back to the definitions
when necessary.

\medskip

\begin{definition}
\label{def:classes}
\begin{enumerate}
\item A {\em Bernoulli sum} is a distribution that 
can be obtained from the sum of finitely many independent 
Bernoulli random variables with possibly different parameters.
The class of Bernoulli sums 
with mean $\lambda$ is denoted by $Ber(\lambda)$ 
and the the union $\cup_{\mu\leq\lambda}Ber(\lambda)$ 
is denoted by $Ber^{\leq}(\lambda).$

\item A distribution $P$ satisfying the inequality
\begin{equation}
\log\frac{P(j)  }{\Po(\lambda,j)  }%
\geq\frac{1}{2}\log\frac{P(j-1)  }{\Po(\lambda,j-1)  }
+\frac{1}{2}\log\frac{P(j+1)  }%
{\Po(\lambda,j+1)  }.\label{eq:ulc}%
\end{equation}
is said to be \emph{ultra log-concave} (ULC); 
cf.\ \cite{Johnson2007a}. The set of
ultra log-concave distributions with mean $\lambda$ shall be denoted
$ULC(\lambda)$, and we also write $ULC^{\leq}(\lambda)$ for the union 
$\cup_{\mu\leq\lambda}ULC(\lambda).$ Note
that (\ref{eq:ulc}) is satisfied for a single value of $\lambda>0$ if
and only if it is satisfied for all $\lambda>0$.

\item The distribution of a random variable $X$ that 
satisfies $E[  X^{\underline{k+1}}]
\leq\lambda E[  X^{\underline{k}}]  $ for all $k\geq0$ will be said
to be \emph{ultra bounded} (UB) with ratio $\lambda$. 
The set of ultra bounded distributions with this
ratio is denoted $UB(\lambda)  .$

\item The distribution of a random variable $X$
satisfying $E[  X^{\underline{k}}]
\leq\lambda^{k}$ for all $k\geq0$ will be said to be \emph{Poisson bounded}
(PB) with ratio $\lambda$.
The set of Poisson bounded distributions with this ratio is denoted 
$PB(\lambda).$

\item A random variable will be said to be 
ULC, UB or PB,
if its distribution is 
ULC, UB or PB, respectively.
\end{enumerate}
\end{definition}

\medskip

First we mention some simple relationships between these classes. Walkup
\cite{walkup} showed that if $X\sim P\in ULC(\lambda)$ 
and $Y\sim Q\in
ULC(\mu)$ then $X+Y\sim P*Q \in 
ULC(\lambda+\mu)$. Hence
$Ber(\lambda) \subseteq ULC(\lambda)$. In \cite{Johnson2007a} it was shown
that, if $P \in ULC(\lambda),  $ then $T_{\alpha}P
\in ULC(  \alpha\lambda)  $. Clearly, $UB(\lambda) \subseteq
PB(\lambda)$. Further, $P$ is Poisson bounded if and only if the 
$\alpha$-thinning
$T_{\alpha} P$ is Poisson bounded, for some $\alpha>0$. The same holds for
ultra boundedness.

\medskip

\begin{proposition}
\label{prop:ub}
In the notation of Definition~\ref{def:classes}, 
$ULC(\lambda)  \subseteq UB(\lambda)  $. That is, if the 
distribution of $X$ is in $ULC(\lambda)$ 
then $E[  X^{\underline{k+1}}]  \leq\lambda
E[  X^{\underline{k}}]  .$
\end{proposition}

\medskip

The next result states that the PB and UB properties
are preserved on summing and thinning.

\medskip

\begin{proposition} \mbox{ } 
\label{prop:preserve}
\begin{enumerate}
\item[(a)] If $X\sim P\in PB(\lambda)$ and $Y\sim Q\in PB(\mu)$ 
are independent, then $X+Y\sim P*Q \in PB(\lambda+\mu)$ 
and $T_\alpha P\in PB(\alpha\lambda).$
\item[(b)] If $X\sim P\in UB(\lambda)$ and $Y\sim Q\in UB(\mu)$ are
independent, then $X+Y\sim P*Q\in UB(\lambda+\mu)$ and 
$T_\alpha P \in UB(\alpha\lambda).$
\end{enumerate}
\end{proposition}

\medskip

Formally, the above discussion can be summarized as,
\[
Ber^{\leq}(  \lambda)  \subseteq ULC^{\leq}(  \lambda)
\subseteq UB(  \lambda)  \subseteq PB(  \lambda)  .
\]
Finally, we note that each of these classes of distributions is
``thinning-convex,'' i.e.,
if $P$ and $Q$ are element of a set then $T_{\alpha}(
P)  \ast T_{1-\alpha}(Q)$ is also an element of 
the same set. In
particular, thinning maps each of these sets into itself,
since 
$T_{\alpha}(P)  =T_{\alpha}(P)  \ast T_{1-\alpha}(\delta_{0})$ 
where $\delta_{0}$, the point mass at zero, has
$\delta_{0}\in Ber^{\leq}(\lambda).$

\newpage

\section{Laws of Thin Numbers: The i.i.d.\ Case\label{sec:ltn}}

In this section we state and prove three versions of the
law of thin numbers,
under appropriate conditions; recall the relevant 
discussion in the Introduction.
Theorem~\ref{thm:ltnweak} proves convergence in total
variation, Theorem~\ref{thm:ltnthermo} in entropy,
and~Theorem~\ref{thm:ltnstrong} in
information divergence.

Recall that the total variation distance $\Vert P-Q\Vert$ 
between two probability distributions $P,Q$ on 
$\mathbb{N}_{0}$ is,
\begin{eqnarray}
\Vert P-Q\Vert:=\sup_{B\subset\mathbb{N}_{0}}|P(B)-Q(B)|=\frac{1}{2}%
\sum_{k\geq0}|P(k)-Q(k)|.
\label{eq:TV}
\end{eqnarray}

\medskip

\begin{theorem}
[weak version]\label{thm:ltnweak} 
For any distribution $P$ on $\mathbb{N}%
_{0}$ with mean $\lambda$, 
\[
\Vert T_{1/n}(P^{\ast n})-\Po(\lambda)\Vert\rightarrow
0,\;\;\;\;n\rightarrow\infty.
\]
\end{theorem}

\medskip

\begin{IEEEproof}
In view of Scheff\'{e}'s lemma, pointwise convergence of discrete
distributions is equivalent to convergence in total variation,
so it suffices to show that,
$T_{1/n}(P^{\ast n})(z)$ converges to $e^{-\lambda}\lambda^z/z!,$
for all $z\geq 0.$

Note that $T_{1/n}(P^{\ast n}) = (  T_{1/n}\left(P)
\right)  ^{\ast n},$ and that (\ref{eq:intro2}) implies the following
elementary bounds for all $\alpha$, using Jensen's inequality:%
\begin{eqnarray}
T_{\alpha}(  P)  (  0)   &  = & \sum_{x=0}^{\infty}P(x)  
(  1-\alpha)^x\geq(  1-\alpha)  ^{\lambda
} \label{eq:tap0} \\
T_{\alpha}(  P)  (  1)   &  = &\sum_{x=1}^{\infty}P(x)  
x\alpha(  1-\alpha)  ^{x-1}. \nonumber%
\end{eqnarray}
Since for i.i.d.\ variables $Y_{i}$, the probability 
$\Pr\{Y_{1}+\ldots+Y_{n}=z\}\geq\binom{n}%
{z}\Pr\{Y_{1}=1\}^{z}P\{Y_{1}=0\}^{n-z}$, 
taking $\alpha=1/n$ we obtain,
\begin{align*}
(  T_{1/n}(  P)  )  ^{\ast n}(  z)   &
\geq\binom{n}{z}\left(  \sum_{x=1}^{\infty}P(x)  \frac{x}%
{n}\left(  1-\frac{1}{n}\right)  ^{x-1}\right)  ^{z}\left(  \left(  1-\frac
{1}{n}\right)  ^{\lambda}\right)  ^{n-z}\\
&  =\frac{n^{\underline{z}}}{n^{z}\, z!}\left(  \sum_{x=1}^{\infty}P(
x)  x\left(  1-\frac{1}{n}\right)  ^{x-1}\right)  ^{z}\left(  1-\frac
{1}{n}\right)  ^{(n-z)  \lambda}.
\end{align*}
Now, for any fixed value of $z$ and $n$ tending to infinity,%
\[
\frac{n^{\underline{z}}}{n^{z}\, z!}\rightarrow\frac{1}{z!},
\]
and%
\[
\left(  1-\frac{1}{n}\right)  ^{(n-z)  \lambda}\rightarrow
e^{-\lambda},
\]
and by monotone convergence,
\[
\sum_{x=1}^{\infty}P(x)  x\left(  1-\frac{1}{n}\right)
^{x-1}\rightarrow\lambda.
\]
Therefore,%
\[
\liminf_{n\rightarrow\infty}(T_{1/n}(P))  ^{\ast
n}(z)  \geq\Po(\lambda,z).
\]
Since all $(T_{1/n}(P))^{\ast n}$ are probability
mass functions
and so is Po$(\lambda)$, the
above $\lim\inf$ is necessarily a limit.
\end{IEEEproof}

\medskip

As usual, the entropy of a probability distribution
$P$ on ${\mathbb N}_0$ is defined by,
$$H(P)=-\sum_{k\geq 0}P(k)\log P(k).$$

\medskip

\begin{theorem}
[thermodynamic version]\label{thm:ltnthermo} For any
Poisson bounded distribution $P$ on 
$\mathbb{N}_{0}$ with mean $\lambda$,
\[
H(  T_{1/n}(  P^{\ast n}))  \rightarrow 
H(\Po(\lambda))  ,\;\;\;\ \text{as }%
n\rightarrow\infty.
\]
\end{theorem}

\medskip

\begin{IEEEproof}
The distribution $T_{1/n}\left(  P^{\ast n}\right)  $ converges pointwise to
the Poisson distribution so, by dominated convergence, it is 
sufficient to prove that $-T_{1/n}\left(
P^{\ast n}\right)  \left(  x\right)  \log\left(  T_{1/n}\left(  P^{\ast
n}\right)  \left(  x\right)  \right)  $ is dominated by a 
summable function. This easily follows from the simple bound
in the following lemma.
\end{IEEEproof}

\medskip

\begin{lemma}
\label{lem:pbb}
Suppose $P$ is Poisson bounded with ratio $\mu$.
Then, $P(x)  \leq \Po\left(  \mu,x\right)  \cdot e^{\mu},$
for all $x\geq 0$.
\end{lemma}

\medskip

\begin{IEEEproof}
Note that, for all $x$, 
\[
P(x)  x^{\underline{k}}\leq\sum_{x=0}^{\infty}P\left(  x\right)
x^{\underline{k}}\leq\mu^{k},
\]
so that, in particular,
$P(x) x^{\underline{x}}\leq\mu^{x}$,
and, $P(x)\leq\frac{\mu^{x}}{x!}=\Po(  \mu,x)\,e^{\mu}.$
\end{IEEEproof}

\medskip

According to \cite[Proof of Theorem 2.5]{Johnson2007a}, 
$H\left(  T_{1/n}\left(  P^{\ast
n}\right)  \right)  \leq H\left( \Po\left(  \lambda\right)  \right)  $ if $P$
is ultra log-concave, so for such distributions the theorem states that the
entropy converges to its maximum. For ultra log-concave distributions the
thermodynamic version also implies convergence in information divergence. 
This also holds for Poisson bounded distributions, which is easily proved 
using dominated convergence. 
As shown in the next theorem, convergence in information
divergence can be established under quite general conditions.

\medskip

\begin{theorem}
[strong version]
\label{thm:ltnstrong} For any distribution
$P$ on
$\mathbb{N}_{0}$ with mean $\lambda$ and 
$D(P\Vert\Po(\lambda))  <\infty$,
\[
D(T_{1/n}(P^{\ast n})\Vert\Po(\lambda))  
\rightarrow0,\;\;\;\ \text{as }n\rightarrow\infty.
\]
\end{theorem}

\medskip

The proof of Theorem~\ref{thm:ltnstrong} is
given in the Appendix; it is based on
a straightforward but somewhat technical
application of the following
general bound.

\medskip

\begin{proposition}
\label{prop:LlogL}
Let $X$ be a random variable with 
distribution $P$ on $\mathbb{N}_{0}$ and
with finite mean $\lambda/\alpha$, 
for some $\alpha\in(0,1)$. If
$D(P\|\Po(\lambda/\alpha))<\infty$,
then,
\be
D(T_{\alpha}(P)\Vert\Po(\lambda))  
\leq
	\frac{\alpha^{2}}{2\left(  1-\alpha\right)  }
	+
	E\left[\alpha X\log\left(\frac{\alpha X}{\lambda}\right)\right]
	<\infty.
\label{eq:pre-var-state}
\ee
\end{proposition}

\medskip

\begin{IEEEproof}
First note that, since $P$ has finite mean,
its entropy is bounded by the entropy of
a geometric with the same mean, which is finite,
so $H(P)$ is finite. Therefore,
the divergence
$D(P\|\Po(\lambda/\alpha))$
can be expanded as,
\be
D(P\|\Po(\lambda))
&=&
	E\left[
	\log\left(\frac{P(X)}{\Po(\lambda/\alpha,X)}\right)
	\right]
	\nonumber\\
&=&
	E[\log(X!)]+\frac{\lambda}{\alpha}-H(P)-
	\frac{\lambda}{\alpha}
	\log\left(\frac{\lambda}{\alpha}\right).
	\nonumber\\
&\geq&
	\frac{1}{2}E[\log^+(2\pi X)]+E[X\log X]
	-H(P)
	-\frac{\lambda}{\alpha}
	\log\left(\frac{\lambda}{\alpha}\right),
	\label{eq:Dexpand}
\ee
where the last inequality follows from 
the Stirling bound,
$$\log(x!)\geq\frac{1}{2}\log^+(2\pi x)+x\log x -x,$$
and $\log^+(x)$ denotes the function $\log\max\{x,1\}$.
Since $D(P\|\Po(\lambda))<\infty$, (\ref{eq:Dexpand})
implies that $E[X\log X]$ is finite. [Recall the
convention that $0\log 0=0$.]

Also note that the representation of $T_\alpha(P)$ in (\ref{eq:intro2}) 
can be written as,
$$T_{\alpha}P(z)  =\sum_{x=0}^{\infty}P(x)  
\Pr\{\mbox{Bin}(x,\alpha)=z\}.
$$
Using this 
and the joint convexity of 
information divergence in its two arguments
(see, e.g.,
\cite[Theorem~2.7.2]{CovThom}),
the divergence of interest can be bounded as,
\begin{align}
D(  T_{\alpha}(P)  \Vert\Po(\lambda))   
&  =D\left(  \sum_{x=0}^{\infty}P(x)  \mathrm{{Bin}%
}(x,\alpha)  \left\Vert \sum_{x=0}^{\infty}P(x)\mathrm{{Po}%
}(\lambda)  \right.  \!\right) \nonumber\\
&  \leq\sum_{x=0}^{\infty}P(x)  D(  \mathrm{{Bin}}(x,\alpha)  
\Vert\Po(\lambda)),
\label{eq:pre-var1}%
\end{align}
where the first term (corresponding 
to $x=0$) equals $\lambda$.
Since the Poisson measures form
an exponential family, they satisfy a
Pythagorean identity 
\cite{csiszar-shields:FnT}
which,
together with the bound,
\be
D(\mathrm{{Bin}}(x,p)\Vert\mathrm{{Po}
}(xp))  \leq\frac{p^{2}}{2(1-p)},
\label{eq:elementary}
\ee
see, e.g.,
\cite{HarRuz} or \cite{KonHarJoh03},
gives, for each $x\geq 1$,
\be
D(\mathrm{{Bin}}(x,\alpha)  \Vert\Po(\lambda))   
&=&
	D(\mathrm{{Bin}}(x,\alpha)
	\Vert\Po(\alpha x))  +D(\mathrm{{Po}}(\alpha x)  
	\Vert\Po(\lambda))
	\nonumber\\
&\leq&
	\frac{\alpha^{2}}{2(1-\alpha)  }+\sum_{j=0}^{\infty
	}\Po(\alpha x,j)  \log\left(  
	\frac{(  \alpha x)^{j}\exp(-\alpha x)/j!}
	{\lambda^{j}\exp(-\lambda)  /j!}\right) 
	\nonumber\\
&=&
	\frac{\alpha^{2}}{2\left(  1-\alpha\right)  }
	+
	\left(  \alpha
	x\log\left(  \frac{\alpha x}{\lambda}\right)  
	-\alpha x+\lambda\right).
	\nonumber
\ee
Since the final bound clearly 
remains valid for $x=0$,
substituting it into (\ref{eq:pre-var1})
gives (\ref{eq:pre-var-state}).
\end{IEEEproof}

\newpage

\section{Laws of Thin Numbers: The Non-i.i.d.\ Case\label{sec:ltn2}}

In this section we state and prove more general
versions of the law of thin numbers,
for sequences of random variables that 
are not necessarily independent 
or identically distributed.
Although some of the results in this section are
strict generalizations of Theorems~\ref{thm:ltnweak}
and~\ref{thm:ltnstrong}, their proofs are different.

We begin by showing that, using a general proof 
technique introduced in \cite{KonHarJoh03}, the 
weak law of thin numbers 
can be established under 
weaker conditions than those in
Theorem~\ref{thm:ltnweak}.
The main idea is to use 
the data-processing inequality
on the total variation 
distance between an appropriate
pair of distributions.

\medskip

\begin{theorem}
[weak version, non-i.i.d.]
\label{thm:ltnweak2} 
Let $P_1,P_2,\ldots$ be an arbitrary sequence
of distributions on ${\mathbb N}_0$,
and write $P^{(n)}=P_1*P_2*\cdots*P_n$ for
the convolution of the first $n$ of them.
Then,
\[
\Vert T_{1/n}(P^{(n)})-\Po(\lambda)\Vert\rightarrow
0,\;\;\;\;n\rightarrow\infty,
\]
as long as the following three conditions
are satisfied as $n\to\infty$:
\begin{itemize}
\item[(a)] $a_n=\max_{1\leq i\leq n}\left[1-T_{1/n}P_i(0)\right]\to 0$;
\item[(b)] $b_n=\sum_{i=1}^n\left[1-T_{1/n}P_i(0)\right]\to\lambda$;
\item[(c)] $c_n=\sum_{i=1}^n\left[1-T_{1/n}P_i(0)-T_{1/n}P_i(1)\right] \to 0$.
\end{itemize}
\end{theorem}

\medskip

Note that Theorem~\ref{thm:ltnweak2} can be 
viewed as a one-dimensional version of
Grigelionis' Theorem~1 
in \cite{grigelionis:63};
recall the relevant comments in the
Introduction. Recently, Schuhmacher
\cite{schuhmacher:05} established
nonasymptotic, quantitative versions
of this result, in terms of the
Barbour-Brown distance, which metrizes
weak convergence in the space of 
probability measures of point processes.
As the information divergence
is a finer functional than the
Barbour-Brown distance, Schuhmacher's
results are not directly comparable
with the finite-$n$ bounds we obtain 
in Propositions~\ref{prop:LlogL},~\ref{prop:var}
and Corollary~\ref{cor:ub}.

Before giving the proof of the theorem, 
we state a simple lemma on a well-known 
bound for $\Vert \po(\lambda)- \po(\mu)\Vert$. 
Its short proof is included for completeness.

\medskip

\begin{lemma}
\label{lem:simpbd} 
\noindent 
For any $\lambda,\mu>0$, 
\[
\Vert\Po(\lambda)-\Po(\mu)\Vert\leq 2\Big[1-e^{-|\lambda-\mu
|}\Big]\leq 2|\lambda-\mu|.
\]
\end{lemma}

\medskip

\begin{IEEEproof}
\noindent Suppose, without loss of generality, that $\lambda>\mu$, 
and define two independent random variables $X\sim$ Po$(\mu)$ 
and $Z\sim$ Po$(\lambda-\mu)$,
so that, $Y=X+Z\sim$ Po$(\lambda)$. 
Then, by the coupling inequality \cite{lindvall:book},
\[
\Vert\Po(\lambda)-\Po(\mu)\Vert\leq 2\Pr\{X\neq
Y\}=2\Pr\{Z\neq0\}=2[1-e^{-(\lambda-\mu)}].
\]
The second inequality in the lemma is trivial.
\end{IEEEproof}

\medskip

{\em Proof of Theorem~\ref{thm:ltnweak2}:}
First we introduce some convenient notation.
Let $X_1,X_2,\ldots$ be independent random variables
with $X_i\sim P_i$ for all $i$; for each $n\geq 1$,
let $Y^{(n)}_1,Y^{(n)}_2,\ldots$ be independent 
random variables with $Y^{(n)}_i\sim T_{1/n}P_i$
for all $i$; and similarly
let $Z^{(n)}_1,Z^{(n)}_2,\ldots$ be independent 
Po($\lambda_i^{(n)})$ random variables, 
where 
$\lambda_i^{(n)} =T_{1/n}P_i(1)$,
for $i,n\geq 1$.
Also we define the sums, 
$S_n=\sum_{i=1}^n Y_i^{(n)}$ 
and $T_n=\sum_{i=1}^n Z_i^{(n)}$,
and note that,
$S_n\sim P^{(n)}$,
and $T_n\sim$ Po($\lambda^{(n)}$), 
where $\lambda^{(n)}=\sum_{i=1}^n\lambda_i^{(n)}$,
for all $n\geq 1$.

Note that $\lambda^{(n)}\to\lambda$ as $n\to\infty$,
since,
$$
\lambda^{(n)}=\sum_{i=1}^n\lambda_i^{(n)}=b_n-c_n,$$
and, by assumption, $b_n\to\lambda$ and $c_n\to 0$,
as $n\to\infty$.

With these definitions in place, 
we approximate,
\be
\|T_{1/n}(P^{(n)})-\mbox{Po}(\lambda)\|
\leq
\|T_{1/n}(P^{(n)})-\mbox{Po}(\lambda^{(n)})\|
+\|\mbox{Po}(\lambda^{(n)})-\mbox{Po}(\lambda)\|,
\label{eq:tria}
\ee
where, by Lemma~\ref{lem:simpbd},
the second term is bounded by
$2|\lambda^{(n)}-\lambda|$ which 
vanishes as $n\to\infty.$
Therefore, it suffices to show
that the first term in (\ref{eq:tria})
goes to zero. For that term,
\ben
\|T_{1/n}(P^{(n)})-\Po(\lambda^{(n)})\|
&=&
	\|P_{S_n}-P_{T_n}\|\\
&\leq&
	\|P_{\{Y_i^{(n)}\}}-
	P_{\{Z_i^{(n)}\}}
	\|\\
&\leq&
	\sum_{i=1}^n
	\|T_{1/n}P_i-
	\mbox{Po}(\lambda_i^{(n)})\|\\
&\leq&
	\sum_{i=1}^n\Big[
	\|T_{1/n}P_i-
	\mbox{Bern}(\lambda_i^{(n)})\|
	+\|\mbox{Bern}(\lambda_i^{(n)})-
	\mbox{Po}(\lambda_i^{(n)})\|\Big],
\een
where the first inequality above follows from
the fact that, being an $f$-divergence, 
the total variation distance satisfies
the data-processing inequality
\cite{csiszar-shields:FnT};
the second inequality comes from the
well-known bound on the total variation
distance between two product measures
as the sum of the distances between
their respective marginals; and
the third bound is simply the triangle
inequality.

Finally, noting that, for any random variable $X\sim P$,
$\|P-\mbox{Bern}(P(1))\|=\Pr\{X\geq 2\}$,
and also recalling the simple estimate,
$$\|\mbox{Bern}(p)-\mbox{Po}(p)\|
=p(1-e^{-p})\leq p^2,$$
yields,
\ben
\|T_{1/n}(P^{(n)})-\Po(\lambda^{(n)})\|
\leq
	c_n
	+\sum_{i=1}^n(\lambda_i^{(n)})^2
\leq
	c_n+\lambda^{(n)}\max_{1\leq i\leq n}
	\lambda_i^{(n)}
\leq	
	c_n+\lambda^{(n)}a_n,
\een
and, by assumption, this converges to zero
as $n\to\infty$, completing the proof.
{\hspace*{\fill}~\IEEEQED\par\endtrivlist\unskip}

\medskip

Recall that, in the i.i.d.\ case,
the weak law of thin numbers only required 
the first moment of $P$ to be finite, 
while the strong version also required that 
the divergence from $P$ to the Poisson 
distribution be finite. For a sum of
independent, non-identically distributed
random variables with finite {\em second}
moments, Proposition~\ref{prop:LlogL} 
can be used as in the proof of 
Theorem~\ref{thm:ltnstrong} to prove the following
result. 
Note that 
the precise conditions required are
somewhat analogous to those 
in Theorem~\ref{thm:ltnweak2}.

\medskip

\begin{theorem}
[strong version, non-i.i.d.]
\label{thm:ltnstrong2} 
Let $P_1,P_2,\ldots$ be an arbitrary sequence
of distributions on ${\mathbb N}_0$,
where each $P_1$ has finite mean $\lambda_i$
and finite variance.
Writing $P^{(n)}$
for
the convolution $P_1*P_2*\cdots*P_n$, 
we have,
\[
D\Big(T_{1/n}(P^{(n)})\Big\|\Po(\lambda)\Big)\rightarrow
0,\;\;\;\;n\rightarrow\infty,
\]
as long as the following two conditions
are satisfied: 
\begin{itemize}
\item[(a)] $\lambda^{(n)}=\frac{1}{n}\sum_{i=1}^n\lambda_i\to\lambda$,
	as $n\to\infty$;
\item[(b)] $\sum_{i=1}^\infty\frac{1}{i^2}E(X^2_i) <\infty$.
\end{itemize}
\end{theorem}

\medskip

The proof of Theorem~\ref{thm:ltnstrong2}
is given in the Appendix, and it is based
on Proposition~\ref{prop:LlogL}.
It turns out that under the additional 
condition of finite second moments, 
the proof of Proposition~\ref{prop:LlogL} 
can be refined to produce a stronger 
upper bound on the divergence.

\medskip

\begin{proposition}
\label{prop:var}
If $P$ is a distribution on $\mathbb{N}_{0}$ with mean
$\lambda/\alpha$ and variance $\sigma^2<\infty$,
for some $\alpha\in(0,1)$, then, 
\begin{equation}
D(T_{\alpha}(P)\Vert\Po(\lambda))  
\leq\alpha^{2}\left(  \frac{1}{2(1-\alpha)  }%
+\frac{\sigma^2}{\lambda}\right)  . 
\label{VarUlig}%
\end{equation}
\end{proposition}

\medskip

\begin{IEEEproof}
Recall that in the proof of Proposition~\ref{prop:LlogL}
it was shown that,
\be
D(  T_{\alpha}(P)  \Vert\Po(\lambda))   
\leq\sum_{x=0}^{\infty}P(x)  D(  \mathrm{{Bin}}(x,\alpha)  
\Vert\Po(\lambda)),
\label{eq:pre-var1b}
\ee
where,
\begin{align}
D(  \mathrm{{Bin}}(x,\alpha)  \Vert\Po(\lambda))   
& \leq \frac{\alpha^{2}}{2\left(1-\alpha\right)  }+\lambda\left(  \frac{\alpha
x}{\lambda}\log\left(  \frac{\alpha x}{\lambda}\right)  -\frac{\alpha
x}{\lambda}+1\right) \nonumber\\
&  \leq\frac{\alpha^{2}}{2\left(  1-\alpha\right)  }+\lambda\left(
\frac{\alpha x}{\lambda}-1\right)  ^{2},
\label{eq:tosub2}%
\end{align}
and where in the last step above we used the simple 
bound $y\log y-y+1\leq y(y-1)-y+1=(y-1)^2,$
for $y>0$.
Substituting (\ref{eq:tosub2}) into (\ref{eq:pre-var1b}) yields,
\begin{align*}
D(T_{\alpha}(P)  \Vert\Po(\lambda))   
&  \leq\sum_{x=0}^{\infty}P(x)\left(
\frac{\alpha^{2}}{2(  1-\alpha)  }+\lambda\left(  \frac{\alpha
x}{\lambda}-1\right)  ^{2}\right) \\
&  =\frac{\alpha^{2}}{2(  1-\alpha)  }+\frac{\alpha^{2}}{\lambda
}\sum_{x=0}^{\infty}P(x)\left(  x-\frac{\lambda}{\alpha
}\right)  ^{2}\\
&  =\frac{\alpha^{2}}{2(  1-\alpha)  }+\frac{\alpha^{2}%
\sigma^2  }{\lambda},
\end{align*}
as claimed.
\end{IEEEproof}

\medskip

Using the bound (\ref{VarUlig})
instead of Proposition~\ref{prop:LlogL},
the following more general version
of the law of thin numbers can be established:

\medskip

\begin{theorem}
[strong version, non-i.i.d.]
\label{thm:noniid} Let $\{X_i\}$ be a sequence 
of (not necessarily independent or identically
distributed) random variables on $\mathbb{N}_{0}$,
and write $P^{(n)}$ for the distribution 
of the partial sum $S_n=X_1+X_2+\cdots+X_n$,
$n\geq 1$. Assume that the $\{X_i\}$
have finite means and variances, and
that:
\begin{enumerate}
\item[(a)]
They are ``uniformly ultra bounded,'' in that,
$\mbox{Var}(X_{i})\leq C E(X_{i})$ for all $i$, 
with a common $C<\infty$;

\item[(b)] Their means satisfy $E(S_n)\to\infty$ as $n\to\infty$;

\item[(c)] Their covariances satisfy,
$$\lim_{n\rightarrow\infty} \frac{\sum_{1\leq i<j\leq n}
\mathrm{{Cov}}(  X_{i},X_{j})  }
{(E(S_n))^2}=0.$$
\end{enumerate}
If in fact $E(X_i)=\lambda>0$ for all $i$, then,
\[
\lim_{n\rightarrow\infty}D(T_{1/n}(P^{(n)}) \Vert 
\Po(\lambda))
=0.
\]
More generally,
\[
\lim_{n\rightarrow\infty}D(T_{\alpha_{n}}(P^{(n)}) \Vert 
\Po(\lambda))
=0,\;\;\;\;\mbox{ where }\alpha_{n}=\lambda/E(S_n).
\]
\end{theorem}

\medskip

\begin{IEEEproof}
Obviously it suffices to prove the general statement.
Proposition~\ref{prop:var} applied to $P^{(n)}$ gives,
\begin{eqnarray*}
D(T_{\alpha_{n}}(P^{(n)}) \Vert \Po(\lambda))   
&\leq&
	\alpha_{n}^{2}
	\left(
	\frac{1}{2(1-\alpha_{n}}+\frac{\mathrm{{Var}}(S_n)
	}{\lambda}\right) \\
&=&
	\frac{\alpha_{n}^{2}}{2(  1-\alpha_{n})  }+\frac{\lambda
	\mathrm{Var}(S_n)}{(E(S_n))^2}\\
&=&
	\frac{\alpha_{n}^{2}}{2(1-\alpha_{n})}
	+
	\frac{\lambda}{(E(S_n))^2}
	\sum_{i=1}^{n}\mathrm{{Var}}(X_{i})  
	+
	\frac{2\lambda}{(E(S_n))^2}
	\sum_{1\leq i<j\leq n}\mathrm{{Cov}}(X_{i},X_{j}).
\end{eqnarray*}
The first and third terms tend to zero by assumptions~(b)
and~(c), respectively. And using assumption~(a), the
second term is bounded above by,
$$	\frac{\lambda}{(E(S_n))^2}\,
	CE(S_n),$$
which
also tends to zero
by assumption~(b). 
\end{IEEEproof}

\newpage

\section{The Thinning Markov Chain} \label{sec:mc}

Before examining the rate of convergence in the law of thin numbers,
we consider a related and somewhat simpler problem for a Markov chain. 
Several of the results in this section may be of independent interest. 
The Markov chain we will discuss was first studied in
\cite[Proof of Theorem~2.5]{Johnson2007a}, and,
within this context, it is a natural discrete analog 
of the Ornstein-Uhlenbeck process associated 
with the Gaussian distribution.

\medskip

\begin{definition}
Let $P$ be a distribution on ${\mathbb N}_0$.
For any  $\alpha\in[0,1]$ and $\lambda>0$, we write
$U_{\alpha}^{\lambda}(P)$ for the 
distribution,
$$U^\lambda_\alpha(P) = T_{\alpha}(P)\ast\mathrm{{Po}%
}((1-\alpha)\lambda).$$
For simplicity, 
$U_{\alpha}^{\lambda}(P)$ is often written
simply as $U_{\alpha}^{\lambda}P$. 
\end{definition}

\medskip

We note that 
$U_{\alpha}^{\lambda}U_{\beta}^{\lambda}=U_{\alpha\beta}%
^{\lambda}$, and that, obviously,
$U_{\alpha}^{\lambda}$ maps probability distributions
to probability distributions. Therefore,
if for a {\em fixed} $\lambda$
we define $Q^t=U^\lambda_{e^{-t}}$ for all $t\geq 0$,
the collection 
$\{Q^t\;;\;t\geq 0\}$
of linear operators 
on the space of probability
measures on ${\mathbb N}_0$ defines
a Markov transition semigroup. Specifically,
for $i,j\in{\mathbb N}_0$, the transition
probabilities,
$$Q^t_{ij}=(Q^t(\delta_i))(j)=
(U^\lambda_{e^{-t}}(\delta_i))(j)
=(T_{e^{-t}}(\delta_i)*\Po((1-e^{-t})\lambda))(j)
=\Pr\{\mbox{Bin}(i,e^{-t})+\Po((1-e^{-t})\lambda)=j\},
$$
define a continuous-time Markov chain 
$\{Z_t\;;\;t\geq 0\}$ on ${\mathbb N}_0$.
It is intuitively clear that, as $\alpha\downarrow 0$
(or, equivalently, $t\to\infty$), the 
distribution $U_\alpha^\lambda P$ should
converge to the $\Po(\lambda)$ distribution.
Indeed, the following two results
state that $\{Z_t\}$ is ergodic, 
with unique invariant measure $\Po(\lambda)$.
Theorem~\ref{thm:chisquare} gives 
the rate at which it converges to $\Po(\lambda)$.

\medskip

\begin{proposition} \label{prop:tvconv}
For any distribution $P$ on ${\mathbb N}_0$, 
$U_{\alpha}^{\lambda}\left(
P\right)$ converges in total variation to 
$\Po\left(  \lambda\right)$, as $\alpha\downarrow 0$.
\end{proposition}

\medskip

\begin{IEEEproof}
From the definition of $U_\alpha^\lambda(P)$, 
\begin{eqnarray}
\Vert U^\lambda_{\alpha}(P)-\Po(\lambda) \Vert
& = &
	\Vert T_{\alpha}(P)\ast\Po((1-\alpha)\lambda)-\Po(\lambda) \Vert
	\nonumber\\
&  = & 
	\Vert (  T_{\alpha}(  P)  -\mathrm{{Po}}(\alpha\lambda))  
	\ast\Po(( 1-\alpha)\lambda)\Vert \nonumber \\
&  \leq & 
	\Vert T_{\alpha}(P)-\Po(\alpha\lambda)\Vert 
	\label{eq:tv1} \\
& = & 
	\frac{1}{2} |(1- T_\alpha (P)(0)) - (1-(\Po(\alpha \lambda,0)) |
	+ \frac{1}{2} \sum_{x=1}^{\infty}
	| T_\alpha (P)(x) - \Po(\alpha \lambda,x) | 
	\nonumber \\
& \leq & 
	\frac{1}{2} [ (1- T_\alpha (P)(0)) + (1-(\Po(\alpha \lambda,0))]
	+ \frac{1}{2} \sum_{x=1}^{\infty}
	( T_\alpha (P)(x) + \Po(\alpha \lambda,x) ) 
	\label{eq:tv2} \\
& = & 
	2-T_{\alpha}(P)(0)-\Po(\alpha\lambda,0), 
	\label{eq:tv3}
\end{eqnarray}
where (\ref{eq:tv1}) follows from the fact that
convolution with any distribution is a contraction with respect
to the $L^1$ norm,
(\ref{eq:tv2}) follows from the triangle inequality, and
(\ref{eq:tv3}) converges to zero because of the bound
(\ref{eq:tap0}).
\end{IEEEproof}

\medskip

Using this, we can give a characterization of the Poisson distribution.

\medskip

\begin{corollary} \label{cor:tvdconv}
Let $P$ denote a discrete distribution with mean $\lambda$. 
If $P= U_{\alpha}^{\lambda}(P)$ for some $\alpha\in(0,1)$,
then $P=\Po(\lambda).$ 
That is, $\Po(\lambda)$ is the unique invariant
measure of the Markov chain $\{Z_t\}$, and,
moreover,
$$D(U_{\alpha}^{\lambda}(P)\Vert
\Po(\lambda))\rightarrow0,\;\;\;\;\mbox{as}\;\alpha\downarrow 0,$$
if and only if $D(U_{\alpha}^{\lambda}(P)\Vert\Po(\lambda))<\infty$ 
for some $\alpha>0.$
\end{corollary}

\medskip

\begin{IEEEproof}
Assume that $P=U_{\alpha}^{\lambda}\left(  P\right)$. Then for any $n$,
$P=U_{\alpha^{n}}^{\lambda}\left(  P\right)$, so for any $\epsilon > 0$,
by Proposition~\ref{prop:tvconv},
$\| P - \Po(\lambda) \| = \| U_{\alpha^{n}}^{\lambda}\left(  P\right) 
- \Po(\lambda) \|
\leq \epsilon$ for $n$ sufficiently large.
The strengthened convergence of 
$D(U_{\alpha}^{\lambda}(P)\Vert \Po(\lambda))$
to zero if $D(U_{\alpha}^{\lambda}(P)\Vert\Po(\lambda))<\infty$ 
can be proved using standard arguments along the
lines of the corresponding discrete-time results
in \cite{Fritz73}\cite{Barron00}\cite{harremoes-holst:09}.
\end{IEEEproof}

\medskip

Next we shall study the rate of convergence 
of $U_{\alpha}^{\lambda}\left(P\right)$ to the Poisson 
distribution. It is easy to check that the
Markov chain $\{Z_t\}$ is in fact {\em reversible}
with respect to its invariant measure $\Po(\lambda)$.
Therefore, the natural setting for the study of
its convergence is the $L^2$ space of functions
$f:{\mathbb N}_0\to{\mathbb R}$ such that,
$E[f(Z)^2]<\infty$ for $Z\sim\Po(\lambda)$.
This space is also endowed with the usual
inner product,
$$\langle f, g\rangle =E[  f(Z)g(Z) ],
\;\;\;\;\mbox{for}\;Z\sim\mbox{Po}(\lambda),
\;f,g\in L^2,
$$
and the linear operators
$U^\lambda_\alpha$ act on functions
$f\in L^2$ by mapping each $f$ into,
$$
(U_\alpha^\lambda f)(x)=E[f(Z_{\alpha,\lambda,x})]
\;\;\;\;\mbox{for}\;Z_{\alpha,\lambda,x}\sim U_\alpha^\lambda(\delta_x).
$$
In other words, 
$$(U_\alpha^\lambda f)(x)=E[Z_{\log(1/\alpha)}|Z_0=x],
\;\;\;\;x\in{\mathbb N}_0.$$
The reversibility of $\{Z_t\}$ with respect
to $\Po(\lambda)$ implies that 
$U_\alpha^\lambda$ is 
a self-adjoint linear operator on $L^2$, 
therefore, its eigenvectors are orthogonal functions. 
In this context, we introduce the
Poisson-Charlier family of orthogonal
polynomials $P_{k}^{\lambda}:$ 

\medskip

\begin{definition}
For given $\lambda$, {\em the Poisson-Charlier polynomial 
of order $k$} is given by,
\[
P_{k}^{\lambda}(x)  
=\frac{1}{(\lambda^{k}k!)^{1/2}}
\sum_{\ell=0}^{k}(-\lambda)^{k-\ell}\binom{k}{\ell}
x^{\underline{\ell}}.
\]
\end{definition}

\medskip

Some well-known properties of 
the Poisson-Charlier polynomials are 
listed in the following lemma
without proof. Note that
their exact form depends on the chosen
normalization; other authors present similar 
results, but with different normalizations.

\medskip

\begin{lemma} For any $\lambda, \mu, k$ and $\ell$:%
\begin{eqnarray}
&\mbox{1)}&
	\hspace{1in}
	\langle P_{k}^{\lambda}, P_{\ell}^{\lambda}\rangle
	=\delta_{k\ell}
	\label{poisnorm}\\
&\mbox{2)}&
	\hspace{1in}
	P_{k+1}^{\lambda}(x)  = \frac{xP_{k}^{\lambda}(x-1)
	-\lambda P_{k}^{\lambda}(x)}{(\lambda(k+1)
	)^{1/2}}
	\label{eq:pcrecrel} \\
&\mbox{3)}&
	\hspace{1in}
	P_k^{\lambda}(x+1) - P_k^{\lambda}(x) =  \Big(\frac{k}{\lambda}
	\Big)^{1/2}
	P_{k-1}^{\lambda}(x) 
	\label{eq:pcrecre2}\\
&\mbox{4)}&
	\hspace{1in}
	P_k^{\lambda + \mu}(x+y) = \sum_{\ell=0}^k 
	\Big(\binom{k}{\ell} \alpha^\ell
	(1-\alpha)^{k-\ell} \Big)^{1/2}
	P_\ell^{\lambda}(x) P_{k-\ell}^{\mu}(y),
	\label{eq:pcrecre3}\\
&&
	\hspace{1in}
	\mbox{where $\alpha = \lambda/(\lambda+ \mu)$.}
\end{eqnarray}
\end{lemma}

\medskip

Observe that,
since the 
Poisson-Charlier polynomials form an orthonormal
set, any function $f\in L^2$ can be expanded
as,
\be
f(x)
=\sum_{k=0}^{\infty}\langle f,P_k^\lambda\rangle
P_{k}^{\lambda}(x).
\label{eq:PCexpand}
\ee
It will be convenient to be able to translate between
factorial moments and the ``Poisson-Charlier moments,''
$E\left[  P^\lambda_{k}\left(  X\right)  \right]  $. For example,
if $X\sim \Po(\lambda)$, then
taking $\ell=0$ in (\ref{poisnorm}) shows that
$E[ P_{k}^{\lambda}(X)]=0$ for all $k\geq1$.
More generally, the following
proposition shows that the role of the Poisson-Charlier 
moments with respect to the Markov chain 
$\{Z_t\}$ is analogous  to the role played by the 
factorial moments with respect to the pure thinning operation;
cf.\ Lemma~\ref{lem:scalmom}. Its proof,
given in the Appendix,
is similar to that
of Lemma~\ref{lem:scalmom}. 

\medskip

\begin{proposition}
\label{Charlierexpo}Let $X\sim P$ be a random variable with 
mean $\lambda$
and write
$X_{\alpha,\lambda}$ for a random variable with
distribution $U_{\alpha}^{\lambda}(P)$. Then,
\[
 E \left[ P_k^{\lambda}(X_{\alpha,\lambda}) \right] = 
\alpha^{k}E\left[  P_{k}^{\lambda
}\left(  X\right)  \right]  .
\]
\end{proposition}

\medskip

If we replace $\alpha$ by $\exp\left(-t\right)$ and assume
that the thinning Markov chain $\{Z_t\}$ has initial
distribution $Z_0\sim P$ with mean $\lambda$, then, 
Proposition~\ref{Charlierexpo} states that,
$$ 
E [ P_k^{\lambda}(Z_t) ] = 
e^{-kt}E[  P_{k}^{\lambda
}(  Z_0) ],
$$
that is,
the Poisson-Charlier moments of $Z_t$
tend to $0$ like $\exp\left(  -kt\right)  E\left[
P_{k}^{\lambda}\left(  Z_0\right)  \right].$ 
Similarly, expanding any $f\in L^2$
in terms of Poisson-Charlier polynomials,
$f\left(  x\right)
=\sum_{k=0}^{\infty}
\langle f,P_k^\lambda\rangle
P_{k}^{\lambda} \left(  x\right)$, and
using Proposition~\ref{Charlierexpo},
$$E[f(Z_t)]
  =E\left[  \sum_{k=0}^{\infty}
\langle f,P_k^\lambda\rangle
P^\lambda_{k}(Z_t)  \right] 
 =\sum_{k=0}^{\infty}
\exp\left(  -kt\right)  
\langle f,P_k^\lambda\rangle
E\left[  P^\lambda_{k}
\left(X\right)  \right].
$$
Thus, the rate of convergence of $\{Z_t\}$ will be dominated by the term
corresponding to $E\left[  P^\lambda_{\kappa}\left(  X\right)  \right],$ 
where $\kappa$ is the first $k\geq1$ such that 
$E\left[  P^\lambda_{k}\left(  X\right) \right]  \neq0.$

The following proposition (proved in the Appendix)
will be used
in the proof of Theorem~\ref{thm:chisquare} below, 
which shows that this is indeed 
the right rate in terms of the $\chi^2$ distance.
Note that there is no restriction on the mean of
$X\sim P$ in the proposition.

\medskip

\begin{proposition} 
\label{prop:radon}
If $X\sim P$ is Poisson bounded, then the 
the likelihood ratio $P/\Po(\lambda)$ can be expanded
as:
\[
\frac{P(x)}{\Po(\lambda,x)} =
\sum_{k=0}^{\infty}
E[P_{k}^{\lambda}(X)]
P_{k}^{\lambda}(x),
\;\;\;\;x\geq 0.
\]
\end{proposition}

\medskip

Assuming $X\sim P\in PB(\lambda)$, 
combining Propositions~\ref{Charlierexpo} 
and~\ref{prop:radon}, we obtain that,
\begin{eqnarray}
\frac{U_{\alpha}^{\lambda}P(x)}{\Po(\lambda,x)}  
& = & 
	\sum_{k=0}^{\infty} E\left[  P_{k}^{\lambda}\left(  X_{\alpha,\lambda}
	\right)  \right]  P_{k}^{\lambda}(x) \nonumber \\
&  = & 1+\sum_{k=\kappa}^{\infty}\alpha^{k}E\left[  P_{k}^{\lambda}\left(
X\right)  \right]  P_{k}^{\lambda}(x) \nonumber \\
&  = & 1+\alpha^{\kappa}\sum_{k=\kappa}^{\infty}\alpha^{k-\kappa}E\left[
P_{k}^{\lambda}\left(  X\right)  \right]  P_{k}^{\lambda}(x), 
\label{eq:edgeworth}
\end{eqnarray}
where, as before, $\kappa$ denotes the first integer $k\geq1$
such that 
$E\left[  P^\lambda_{k}\left(  X\right) \right]  \neq0.$
This sum can be viewed as a discrete analog
of the well-known Edgeworth expansion for the
distribution of a continuous random variable. A
technical disadvantage
of both this and the standard Edgeworth expansion
is that,
although the sum converges in $L^{2}$, truncating it
to a finite number of terms
in general produces an expression which may 
take negative values. By a more detailed analysis we 
shall see in the following two sections
how to get around this problem. 

For now, we determine the rate of convergence
of $U_\alpha^\lambda P$ to $\Po(\lambda)$ in terms
of the $\chi^2$ distance
between $U_\alpha^\lambda P$ and $\Po(\lambda)$;
recall the definition of the $\chi^2$ distance
between two probability distributions $P$ and
$Q$ on ${\mathbb N}_0$:
$$
\chi^{2}\left(  P,Q\right)  =\sum_{x=0}^{\infty
}Q\left(  x\right)  \left(  \frac{P\left(  x\right)  }{Q\left(  x\right)
}-1\right)  ^{2}.
$$

\medskip

\begin{theorem}
\label{thm:chisquare}
If $X\sim P$ is Poisson bounded,
then 
$\chi^{2}(U_{\alpha}^{\lambda}P,\Po(\lambda))$ is
finite for all $\alpha\in[0,1]$ and,
\[
\frac{\chi^{2}\left(  U_{\alpha}^{\lambda}P,\Po\left(
\lambda\right)  \right)  }{\alpha^{2\kappa}}\rightarrow E\left[  P_{\kappa
}^{\lambda}\left(  X\right)  \right]  ^{2},
\;\;\;\;\mbox{as}\;\alpha\downarrow 0,
\]
where $\kappa$ denotes the smallest $k>0$ such that 
$E\left[  P_k ^{\lambda}\left(  X\right)  \right]  \neq0.$
\end{theorem}

\medskip

\begin{IEEEproof}
The proof is based on a Hilbert space argument 
using the fact that the Poisson-Charlier 
polynomials are orthogonal. 
Suppose $X\sim P\in PB(\mu)$.
Using Proposition~\ref{prop:radon},
\begin{align*}
\chi^{2}(U_{\alpha}^{\lambda}P,\Po(\lambda))    
&
=\sum_{x=0}^{\infty}\Po(\lambda,x)
\left(\frac{U_{\alpha}^\lambda P(x)}{\Po(\lambda,x)}-1\right)^2\\
& 
=\sum_{x=0}^{\infty}\Po(\lambda,x)
\left(\sum_{k=\kappa}^{\infty}\alpha^kE[P_k^\lambda(X)]
P_k^\lambda(x)\right)^2\\
&
=\sum_{k=\kappa}^{\infty}\alpha^{2k}E[P_k^\lambda(X)]^2,
\end{align*}
where the last step follows from the orthogonality relation (22). 
For $\alpha=1$ we have,
\begin{align*}
\chi^2(P,\Po(\lambda))& =\sum_{x=0}^{\infty}\Po(\lambda,x)
\left(\frac{P(x)}{\Po(\lambda,x)}-1\right)^2\\
& =\sum_{x=0}^{\infty}\Po(\lambda,x)
\left(\frac{P(x)}{\Po(\lambda,x)}\right)^2-1,
\end{align*}
which is finite.
From the previous expansion we see that 
$\chi^2(U_\alpha^\lambda P,\Po(\lambda))$ 
is increasing in $\alpha$, which implies the
finiteness claim. Moreover, that expansion
has $\alpha^{2\kappa}E[P_{\kappa}^{\lambda}(X)]^2$
as its dominant term, implying the stated limit.
\end{IEEEproof}

\medskip

Theorem~\ref{thm:chisquare} readily leads to 
upper bounds on the rate of convergence in terms
of information divergence via 
the standard bound,
$$
D(P\Vert Q) \leq \log (1+\chi^{2}(P,Q))
\leq \chi^{2}(P,Q),
$$
which follows from direct applications
of Jensen's inequality.
Furthermore, replacing this bound
by the well-known approximation 
\cite{csiszar-shields:FnT},
\[
D\left(  P\Vert Q\right)  \approx\frac{1}{2}\chi^{2}\left(  P,Q\right),
\]
gives the estimate,
\[
D(U_{\alpha}^{\lambda}P\Vert\Po(\lambda))
\approx\alpha^{2\kappa}\frac{E\left[  P_{\kappa}^{\lambda}\left(
X\right)  \right]  ^{2}}{2}=\frac{E\left[  P_{\kappa}^{\lambda}\left(
U_{\alpha}X\right)  \right]  ^{2}}{2}.
\]
We shall later prove that, in certain cases,
this approximation can indeed
be rigorously justified.

\newpage

\section{The Rate of Convergence in the Strong Law 
of Thin Numbers\label{sec:ub}}

Let $X\sim P$ be a random variable on ${\mathbb N}_0$ with
mean $\lambda$. In Theorem~\ref{thm:ltnstrong} we showed
that, if $D(P\|\Po(\lambda))$ is finite, then,
\be
D(T_{1/n}(P^{*n})\|\Po(\lambda))\to0,
\;\;\;\;\mbox{as}\;n\to\infty.
\label{eq:ltnI}
\ee
If $P$ also has finite variance $\sigma^2$,
then Proposition~\ref{prop:var} implies that,
for all $n\geq 2$,
\begin{equation}
D\left(  T_{1/n}\left(  P^{\ast n}\right)  \Vert\Po\left(
\lambda\right)  \right)  \leq
\frac{\sigma^2}{n\lambda}
+\frac{1}{n^2},
\label{eq:linear}
\end{equation}
suggesting a convergence rate of order $1/n$.
In this section, we prove more precise upper 
bounds on the rate of convergence in the strong 
law of thin numbers (\ref{eq:ltnI}). For example, 
if $X$ is an ultra bounded random variable with 
$\sigma^2\neq\lambda$, then
we show that in fact,
\[
\limsup_{n\rightarrow\infty}n^{2}D\left(  T_{1/n}(  P^{\ast
n})  \Vert\Po(  \lambda)  \right)  \leq2c^{2},
\]
where $c=E[P_2^\lambda(X)]=(\sigma^2-\lambda)/(\lambda\sqrt{2})\neq 0$.
This follows from the more general result
of Corollary~\ref{cor:ub};
its proof is based 
on a detailed analysis of the {\em scaled
Fisher information} introduced in
in \cite{KonHarJoh03}.
We begin by briefly reviewing some properties 
of the scaled Fisher information:

\medskip

\begin{definition}
The {\em scaled Fisher information} of a random variable 
$X\sim P$ with mean $\lambda$,
is defined by,
\[
K(X) =K(P)  =\lambda E\left[  \rho_{X}\left(  X\right)  ^{2}\right]
\]
where $\rho_{X}$ denotes the scaled score function,
\[
\rho_{X}\left(  x\right)  =\frac{\left(  x+1\right)  P\left(  x+1\right)
}{\lambda P\left(  x\right)  }-1.
\]
\end{definition}

\medskip

In \cite[Proposition~2]{KonHarJoh03} it was shown,
using a logarithmic Sobolev inequality 
of Bobkov and Ledoux \cite{bobkov3}, that
for any $X\sim P$,
\be
D\left(  P\Vert\Po\left(  \lambda\right)  \right)  \leq K\left(
X\right),
\label{eq:logS}
\ee
under mild conditions on the support of $P$.
Also, \cite[Proposition~3]{KonHarJoh03} states that
$K(X)$ satisfies a subadditivity property:
For independent random variables $X_{1},X_{2},\ldots,X_{n}$,
\be
K\left(  \sum_{i=1}^{n}X_{i}\right)  \leq\sum_{i=1}^{n}
\frac{E[X_i]}{\lambda} K(X_i)
\label{eq:Ksub} 
\ee
where $\lambda=\sum_i E(X_{i}).$ 
In particular, recalling that the thinning
of a convolution is the convolution of the 
corresponding thinnings,
if $X_{1},X_{2},\ldots,X_{n}$ are i.i.d.\ random
variables with mean $\lambda$ then the bounds
in (\ref{eq:logS}) and (\ref{eq:Ksub})
imply,
\be
D\left(  T_{1/n}(  P^{\ast n})  \Vert\Po\left(
\lambda\right)  \right)  \leq K\left(T_{1/n}(P)\right)  .
\label{eq:oldK}
\ee
Therefore, our next goal is to determine the rate at which 
$K(T_{\alpha}(X))$ tends to $0$ for $\alpha$ tending to $0.$ 
We begin with the following proposition; 
its proof is given in Appendix.

\medskip

\begin{proposition}
\label{prop:altsum} 
If $X\sim P$ is Poisson bounded, then
$P$ admits the 
representation,
\[
P(x)  =\frac{1}{x!}\sum_{\ell=0}^{\infty}\left(  -1\right)
^{\ell}\frac{E\left[X^{\underline{x+\ell}}\right]  }{\ell!}.
\]
Moreover, the truncated sum from $\ell=0$ to $m$
is an upper bound for $P(x)$ if $m$ is
even, and a lower bound if $m$ is odd.
\end{proposition}

\medskip

An important consequence of this proposition is that 
$T_{\alpha}P(x)$ tends to zero like $\alpha^{x}$, 
as $\alpha\downarrow 0.$ 
Moreover, it leads to the following asymptotic
result for the scaled Fisher information,
also proved in the Appendix.

\medskip

\begin{theorem}
\label{thm:kthin} 
Suppose $X\sim P$ has mean $\lambda$ and it is ultra bounded 
with ratio $\lambda$.
Let $\kappa$ denote the smallest integer $k\geq 1$
such that $E\left[ P_{k}^\lambda\left(  X\right)  \right]  \neq 0$.
Then, 
\[
\lim_{\alpha\rightarrow0}\frac{K\left(  T_{\alpha}P\right)  }{\alpha^{\kappa}%
}=\kappa c^{2},
\]
where $c=E\left[  P_{\kappa}^\lambda\left(  X\right)  \right]  .$
\end{theorem}

\medskip

Combining Theorem~\ref{thm:kthin} with (\ref{eq:oldK}) immediately yields:

\medskip

\begin{corollary}
\label{cor:ub} 
Suppose $X\sim P$ has mean $\lambda$ and it is ultra bounded 
with ratio $\lambda$.
Let $\kappa$ denote the smallest integer $k\geq 1$
such that $E\left[ P_{k}^\lambda\left(  X\right)  \right]  \neq 0$.
Then,
\[
\limsup_{n\rightarrow\infty}n^{\kappa}D\left(  T_{1/n}\left(  P^{\ast
n}\right)  \Vert\Po\left(  \lambda\right)  \right)  \leq\kappa
c^{2},
\]
where $c=E\left[  P_{\kappa}^\lambda\left(  X\right)  \right]  .$
\end{corollary}

\newpage

\section{Monotonicity Results for the Scaled Fisher Information} 
\label{sec:monotone}

In this section we establish a finer result for the
behavior of the scaled Fisher information upon 
thinning, and use that to deduce a stronger 
finite-$n$ upper bound for the strong law of thin numbers.
Specifically, if $X\sim P$ is ULC with mean $\lambda$,
and $X_\alpha$ denotes a random variable
with distribution $T_\alpha P$,
we will show that $K(X_\alpha) \leq \alpha^2 K(X)$.
This implies that, for all ULC random variables $X$,
we have the following finite-$n$ version of the strong
law of thin numbers,
$$D(T_{1/n}(P^{*n})\|\Po(\lambda))\leq\frac{K(X)}{n^2}.$$
Note that, unlike the more general result in (\ref{eq:linear})
which gives a bound of order $1/n$, the above bound
is of order $1/n^2$, as long as $X$ is ULC.

The key observation for these results
is in the following lemma.

\medskip

\begin{lemma} 
\label{lem:kder}
Suppose $X$ is a ULC random variable with distribution
$P$ and mean $\lambda$. For any $\alpha\in (0,1)$,
write $X_\alpha$ for a random variable with distribution
$T_\alpha P$. Then the derivative of $K(X_\alpha)/\alpha$ 
with respect to $\alpha$ satisfies,
$$ \frac{\partial}{\partial \alpha} \left( \frac{K(X_\alpha)}{\alpha} \right)
= \frac{1}{\alpha^2} S(X_\alpha),$$
where, for a random variable $Y$ with mass 
function $Q$ and mean $\mu$,
we define,
$$
S(Y) = \sum_{y=0}^{\infty} 
\frac{Q(y+1) (y+1)}{\mu Q(y)} 
\left( \frac{Q(y+1)(y+1)}{Q(y)} - \frac{Q(y+2)(y+2)}{Q(y+1)}
\right)^2.$$
\end{lemma}

\medskip

\begin{IEEEproof}
This result follows on using the expression for the derivative of 
$T_\alpha P$ arising as the case $f(\alpha) = g(\alpha) = 0$ in Proposition
3.6 of \cite{Johnson2007a}, that is,
$$ \frac{\partial}{\partial \alpha} (T_\alpha P)(x) = \frac{1}{\alpha} 
\Big[ x (T_\alpha P)(x) - (x+1) (T_\alpha P)(x) \Big].$$
Using this, for each $x$ we deduce that,
\begin{eqnarray*} 
\lefteqn{ \frac{\partial}{\partial \alpha} \left( 
\frac{((T_\alpha P)(x+1))^2 (x+1)^2}{
\alpha^2 (T_\alpha P)(x) \lambda} \right) } \\
& = & \frac{(T_\alpha P)(x+1) (x+1)}{\alpha^3 \lambda}
\left( \frac{(T_\alpha P)(x+1)(x+1)}{(T_\alpha P)(x)} - \frac{(T_\alpha P)(x+2)(x+2)}{(T_\alpha P)(x+1)}
\right)^2 \\ 
& & + \frac{1}{\alpha^3 \lambda}
\left( \frac{((T_\alpha P)(x+1))^2(x+1)^2 x}{(T_\alpha P)(x)}
- \frac{((T_\alpha P)(x+2))^2(x+2)^2 (x+1)}{(T_\alpha P)(x+1)} \right).
\end{eqnarray*}
The result follows (with the term-by-term differentiation of the infinite
sum justified) if the sum of these terms in $x$ is absolutely convergent. The
first terms are positive, and their sum is absolutely convergent to $S$ by
assumption. The second terms form a collapsing sum,
which is absolutely convergent assuming that,
$$\sum_{x=0}^{\infty} \frac{((T_\alpha P)(x+1))^2(x+1)^2 x}{(T_\alpha P)(x)}
<\infty.$$
Note that, for any ULC distribution $Q$, by definition we have for all $x$,
$(x+1)Q(x+1)/Q(x)\leq xQ(x)/Q(x-1)$, so that the above
sum is bounded above by,
$$\frac{(T_\alpha P)(1)}{(T_\alpha P)(0)} \sum_x (T_\alpha P)(x+1) (x+1)x,$$
which is finite by Proposition~\ref{prop:ub}.
\end{IEEEproof}

\medskip

We now deduce the following theorem, which
parallels Theorem 8 respectively of \cite{yu:thinning:pre}, 
where a corresponding result is proved for the information
divergence.

\medskip

\begin{theorem} 
Let $X\sim P$ be a ULC random variable with mean $\lambda$.
Write $X_\alpha$ for a random variable with distribution
$T_\alpha P$. Then:
\begin{eqnarray}
&(i)&
\hspace{0.5in}
K(X_\alpha) \leq \alpha^2 K(X),
\;\;\;\;\alpha\in(0,1);\\
&(ii)&
\hspace{0.5in}
D(T_{1/n}(P^{*n})\|\Po(\lambda))\leq\frac{K(X)}{n^2},
\;\;\;\;n\geq 2.
\end{eqnarray}
\end{theorem}

\medskip

\begin{IEEEproof} 
The first part follows from the observation that
$K(T_\alpha X)/\alpha^2$ is increasing in $\alpha$,
since, by Lemma~\ref{lem:kder},
its derivative is $(S(T_\alpha X) - K(T_\alpha X))/\alpha^3$.
Taking $g(y) = P(y+1) (y+1)/P(y)$ in the more technical
Lemma~\ref{lem:poincarish} below, we deduce that 
$S(Y) \geq K(Y)$ for any random
variable $Y$, and this proves~$(i)$. 
Then~$(ii)$ immediately follows from~$(i)$ 
combined with the earlier bound~(\ref{eq:oldK}),
upon recalling that thinning preserves the ULC
property \cite{Johnson2007a}.
\end{IEEEproof}

\medskip

Consider the finite 
difference operator $\Delta$ defined
by, $(\Delta g)(x) = g(x+1) - g(x)$, for
functions $g: {\mathbb N}_0\rightarrow {\mathbb R}$.
We require a result suggested by relevant
results in \cite{borovkov2}\cite{klaassen}.
Its proof is given in the Appendix.

\medskip

\begin{lemma} \label{lem:poincarish}
Let $Y$ be ULC random variable with 
distribution $P$ on ${\mathbb N}_0$.
Then for any function $g$, defining $\mu = \sum_y P(y) g(y)$,
$$ \sum_{y=0}^{\infty} 
P(y) (g(y) - \mu)^2 \leq \sum_{y=0}^{\infty} P(y+1) (y+1) \Delta g(y)^2. $$
\end{lemma}

\newpage

\section{Bounds in Total Variation\label{sec:tv}}

In this section, we show that a modified version of the argument used in the
proof of Proposition~\ref{prop:var} gives an upper bound
to the rate of convergence in the weak law of small numbers.
If $X\sim P$ has mean $\lambda$ and variance $\sigma^2$,
then combining the bound (\ref{VarUlig}) 
of Proposition~\ref{prop:var}
with Pinsker's inequality we obtain,
\be
\Vert T_{1/n}\left(  P^{\ast n}\right)  -\Po(\lambda)\Vert
\leq\left(  \frac{1}{2n^{2}\left(  1-n^{-1}\right)  }+\frac{\sigma^2
}{n\lambda}\right)  ^{1/2},
\label{eq:weakB}
\ee
which gives an upper bound of order
$n^{-1/2}.$ From the asymptotic upper bound on information divergence,
Corollary~\ref{cor:ub}, we know that one should be able to obtain upper bounds
of order $n^{-1}.$ Here we derive an upper bound on total variation using
the same technique used in the proof of Proposition~\ref{prop:var}.

\medskip 
\begin{theorem}
\label{2TV} 
Let $P$ be a distribution on $\mathbb{N}_{0}$ with finite
mean $\lambda$ and variance $\sigma^{2}$. Then,
\[
\left\Vert T_{1/n}(P^{\ast n})-\Po(\lambda)\right\Vert \leq\frac
{1}{n2^{1/2}}+\frac{\sigma}{n^{1/2}}\min\left\{  1,\frac{1}{2\lambda^{1/2}%
}\right\}  ,
\]
for all $n\geq2$.
\end{theorem}

\medskip

The proof uses the following simple bound, which follows easily 
from a result of Yannaros, \cite[Theorem~2.3]{Yannaros1991};
the details are omitted.

\medskip

\begin{lemma}
\label{Lemma3}For any $\lambda>0$, $m\geq1$ and $t\in(0,1/2]$, we have,
\[
\Vert\mathrm{{Bin}}(m,t)-\Po(\lambda)\Vert\leq t2^{-1/2}%
+|mt-\lambda|\min\left\{  1,\frac{1}{2\lambda^{1/2}}\right\}  .
\]
\end{lemma}

\medskip

\begin{IEEEproof}
The first inequality in the proof 
of Proposition~\ref{prop:var} remains valid 
due to the convexity
of the total variation norm (since it is an $f$-divergence). 
The next equality
becomes an inequality, and it is justified by the triangle,
and we have:
\begin{align}
\Vert T_{1/n}(P^{*n})-\Po(\lambda )\Vert &
=\frac{1}{2}\sum_{x\geq 0}\left\vert \sum_{y\geq0}P^{*n}(y)
\Big[
\Pr\{\mbox{Bin}(y,1/n)=x\}-\Po(\lambda,x)\Big]
\right\vert \nonumber\\
&  \leq\sum_{y\geq0}P^{*n}(y)\frac{1}{2}\sum_x\left\vert  
\Pr\{\mathrm{{Bin}}(y,1/n)=x\}-\Po(\lambda,x )
\right\vert \nonumber\\
&  =\sum_{y\geq0}P^{*n}(y)\Vert\mathrm{{Bin}}(y,1/n)-\Po(\lambda)
\Vert.\nonumber
\end{align}
And using Lemma~\ref{Lemma3} leads to,
\begin{align*}
\Vert T_{1/n}(P^{\ast n})-\Po(\lambda)\Vert &  \leq\sum_{y\geq
0}P^{\ast n}(y)\Vert\mathrm{{Bin}}(y,1/n)-\Po(\lambda)\Vert.\\
&  =\sum_{y\geq0}P^{\ast n}(y)\left(  \frac{1}{n2^{1/2}}+\left\vert
\frac{y}{n}-\lambda\right\vert \min\left\{  1,\frac{1}{2\lambda^{1/2}%
}\right\}  \right)  ,
\end{align*}
and the result follows by an application
of H\"{o}lder's inequality.
\end{IEEEproof}

\newpage

\section{Compound Thinning\label{sec:comp}}

There is a natural generalization of the thinning operation, 
via a process which closely parallels 
the generalization of the Poisson distribution to the compound
Poisson. Starting with a random variable $Y\sim P$ with values in
$\mathbb{N}_{0}$, the $\alpha$-thinned version of $Y$ 
is obtained by writing $Y=1+1+\cdots+1$ ($Y$ times), 
and then keeping each of these $1$s with
probability $\alpha$, independently of all the others; 
cf.~(\ref{eq:intro}) above.

More generally, we choose and fix a ``compounding''
distribution $Q$ on $\mathbb{N}=\{1,2,\ldots\}$.
Given $Y\sim P$ on ${\mathbb N}_0$ and $\alpha\in[0,1]$,
then the {\em compound $\alpha$-thinned version of 
$Y$ with respect to Q}, or, for
short, the \emph{$(\alpha,Q)$-thinned version of $Y$}, 
is the random variable which results from 
first thinning $Y$ as above and then replacing
of the $1$s that are kept by 
an independent random sample from $Q$,
\begin{equation}
\sum_{n=1}^{Y}B_{n}\xi_{n},\;\;\;\;B_{i}\sim\text{i.i.d.\ Bern}%
(\alpha),\;\xi_{i}\sim\text{i.i.d.\ }Q, \label{eq:thinned-sum}%
\end{equation}
where all the random variables involved are independent.
For fixed $\alpha$ and $Q$, we write
$T_{\alpha,Q}(P)$ for the distribution 
of the $(\alpha,Q)$-thinned version of
$Y\sim P.$ Then $T_{\alpha,Q}(P)$ can be expressed as a mixture of
\textquotedblleft compound binomials\textquotedblright\ in the same way as
$T_{\alpha}(P)$ is a mixture of binomials. The \emph{compound binomial
distribution} with parameters $n,\alpha,Q,$ denoted $\mathrm{{CBin}}%
(n,\alpha,Q)$, is the distribution of the sum of $n$ i.i.d.\ random variables,
each of which is the product of a Bern$(\alpha)$ random variable and an
independent $\xi\sim Q$ random variable. In other words, it is the
$(\alpha,Q)$-thinned version of the point mass at $n$, i.e., the distribution
of (\ref{eq:thinned-sum}) with $Y=n$ w.p.1. Then we can express the
probabilities of the $(\alpha,Q)$-thinned version of $P$ as, 
$T_{\alpha,Q}(P)(k)=
\sum_{\ell\geq k}P(\ell)\Pr\{\mathrm{{CBin}}(\ell,\alpha,Q)=k\}$.

The following two observations are immediate from the definitions.

\begin{enumerate}
\item 
Compound thinning maps a Bernoulli sum into a compound Bernoulli
sum: If $P$ is the distribution of the Bernoulli sum $\sum_{i=1}^{n}B_{i}$
where the $B_{i}$ are independent Bern$(p_{i})$, then $T_{\alpha,Q}(P)$
is the distribution of the ``compound Bernoulli
sum,'' $\sum_{i=1}^{n}B_{i}^{\prime}\xi_{i}$ where the
$B_{i}^{\prime}$ are independent Bern$(\alpha p_{i})$, and the $\xi_{i}$
are i.i.d.\ with distribution $Q$, independent of the $B_{i}$.

\item Compound thinning maps the Poisson to the compound Poisson distribution,
that is,
$T_{\alpha,Q}(\Po(\lambda))=\mbox{CPo}(\alpha\lambda,Q)$,
the compound Poisson distribution
with rate $\alpha\lambda$ and compounding distribution $Q$. 
Recall that $\mbox{CPo}(\lambda,Q)$ is defined
as the distribution of,
\[
\sum_{i=1}^{\Pi_{\lambda}}\xi_{i},
\]
where the $\xi_{i}$ are as before, and $\Pi_{\lambda}$ is a 
$\Po(\lambda)$ random variable that is independent of the $\xi_{i}$.
\end{enumerate}

Perhaps the most natural way in which the compound Poisson distribution arises
is as the limit of compound binomials. That is, 
$\mbox{CBin}(n,\lambda/n,Q)\rightarrow\mbox{CPo}(\lambda,Q),$ as $n\rightarrow
\infty$, or, equivalently, 
\[
T_{1/n,Q}\left(  \mathrm{{Bin}}(n,\lambda)\right)  =T_{1/n,Q}(P^{\ast
n})\rightarrow\mathrm{{CPo}}(\lambda,Q),
\]
where $P$ denotes the Bern$(\lambda)$ distribution. 

As with the strong law of thin numbers, this results
remains true for general distributions $P$, and the
convergence can be established in the sense of
information divergence:

\medskip

\begin{theorem}
\noindent Let $P$ be a distribution on $\mathbb{N}_{0}$ with mean $\lambda>0$
and finite variance $\sigma^{2}$. Then, for any probability measure $Q$ on
$\mathbb{N}$,
\[
D(T_{1/n,Q}(P^{\ast n})\Vert \mathrm{{CPo}}(\lambda,Q))
\rightarrow0,\;\;\;\;\text{as}\ n\rightarrow\infty,
\]
as long as 
$D(P\Vert \mathrm{{Po}}(\lambda))<\infty$.
\end{theorem}

\medskip

The proof is very similar to that of Theorem~\ref{thm:ltnstrong} 
and thus
omitted. In fact, the same argument as that proof works for
non-integer-valued compounding. That is, if $Q$ is an \emph{arbitrary}
probability measure on $\mathbb{R}^{d}$, then compound thinning
a $\mathbb{N}_{0}$-valued random variable $Y\sim P$ 
as in (\ref{eq:thinned-sum}) gives a probability
measure $T_{\alpha,Q}(P)$ on $\mathbb{R}^{d}$.

It is somewhat remarkable that the statement \emph{and} proof of
most of our results concerning the information divergence
remain essentially unchanged in this case. For example,
we easily obtain the following analog of Proposition~\ref{prop:var}.

\medskip

\begin{proposition}
\noindent If $P$ is a distribution on $\mathbb{N}_{0}$ with mean 
$\lambda/\alpha$ and variance $\sigma^2<\infty$, for some
$\alpha\in(0,1)$, then,
for any
probability measure $Q$ on $\mathbb{R}^{d}$,
\[
D( T_{\alpha,Q}(P)  \Vert \Cpo(
\lambda,Q ) )  \leq\alpha^2\left(  \frac{1}{2(1-\alpha)}
+\frac{\sigma^2}{\lambda}\right)  .
\]
\end{proposition}

The details of the argument of the proof
of the proposition are straightforward 
extensions of the corresponding proof of
Proposition~\ref{prop:var}.

\newpage

\section*{Acknowledgement} 

The authors wish to thank Emre Telatar 
and Christophe
Vignat for hosting a small workshop in January 2006,
during which some of these ideas developed. Jan
Swart also provided us with useful comments.

\appendix

{\em Proof of Lemma~\ref{lem:scalmom}:}
Simply apply Lemma~\ref{lem:fallfact} to Definition~\ref{def:thin} with
$Y\sim P$, to obtain,
\ben
E[Y_{\alpha}^{\underline{k}}]  
&=&
	E\Big[  \Big(  \sum_{x=1}^{Y}B_{x}\Big)^{\underline{k}}\Big]\\
&=&
	E\Big\{  E\Big[  \Big(
	\sum_{x=1}^{Y}B_{x}\Big)^{\underline{k}}\,\Big|\,Y\Big]
	\Big\}\\
&=&
	E\Big\{  E\Big[  
	\sum_{k_{x}\in\{0,1\},\;\sum k_x=k}
	k!
	{\displaystyle\prod\limits_{x=1}^{Y}}
	B_x^{\underline{k_x}}
	\,\Big|\,Y\Big]
	\Big\}\\
&=&
	E\Big[\binom{Y}{k} k!\alpha^{k}\Big]\\
&=&
	\alpha^{k}E[Y^{\underline{k}}], 
\een
using the fact that the sequence of 
factorial moments of the Bern($\alpha$)
distribution are
$\{1,\alpha,0,0,\ldots\}$.
{\hspace*{\fill}~\IEEEQED\par\endtrivlist\unskip}

\medskip

{\em Proof of Proposition~\ref{prop:inj}:}
Assume that $T_{\alpha_{0}}P=T_{\alpha_{0}}Q$ for a given
$\alpha_{0}>0.$ Then, recalling the property 
stated in~(\ref{eq:semigroup}), it follows that,
$T_{\alpha}P=T_{\alpha}Q$ for all $\alpha\in[  0,\alpha_{0}]$.
In particular, $T_{\alpha}P(0)  =T_{\alpha}Q(0)$
for all $\alpha\in[0,\alpha_0]$, i.e.,
\[
\sum_{x=0}^{\infty}P(x)  (1-\alpha)^x=\sum
_{x=0}^{\infty}Q(x)(1-\alpha)^x,%
\]
for all $\alpha\in[  0,\alpha_{0}],$ which is only possible if
$P(x)  =Q(x)$ for all $x\geq 0$.
{\hspace*{\fill}~\IEEEQED\par\endtrivlist\unskip}

\medskip

{\em Proof of Proposition~\ref{prop:ub}:}
Note that the expectation,
\[
\sum_{x=0}^{\infty}P(x)  x^{\underline{k}}\Big(  \frac{(
x+1)  P(x+1)  }{\lambda P(x)  }-1\Big)
\leq 0,
\]
by the Chebyshev rearrangement lemma, since it is the
covariance between an increasing and a decreasing function. 
Rearranging this inequality gives,
\[
E[  X^{\underline{k+1}}]  =\sum_{x=0}^{\infty}P(x+1)
(  x+1)  ^{\underline{k+1}}\leq\lambda\sum_{x=0}^{\infty}P(x)  
x^{\underline{k}}=\lambda E[  X^{\underline{k}}],
\]
as required.
{\hspace*{\fill}~\IEEEQED\par\endtrivlist\unskip}

\medskip

{\em Proof of Proposition~\ref{prop:preserve}:}
To prove part (a), using Lemma~\ref{lem:fallfact}, we have,
\begin{align*}
E[(X+Y)^{\underline{k}}]   &  =E\Big[  \sum
_{\ell=0}^{k}\binom{k}{\ell}X^{\underline{\ell}}Y^{\underline{k-\ell}}\Big]  \\
&  =\sum_{\ell=0}^{k}\binom{k}{\ell}E[  X^{\underline{\ell}}]  E[
Y^{\underline{k-\ell}}]  \\
&  \leq\sum_{\ell=0}^{k}\binom{k}{\ell}\lambda^{\ell}\mu^{k-\ell}\\
&  =(  \lambda+\mu)  ^{k}.
\end{align*}
It is straightforward to check, using Lemma~\ref{lem:scalmom},
that $T_\alpha P \in PB(\alpha\lambda)$.

To prove part (b), using Lemma~\ref{lem:fallfact}, 
Pascal's identity and relabelling, yields,
\begin{align*}
E[  (  X+Y)  ^{\underline{k+1}}]   &  =E\left[
\sum_{\ell=0}^{k+1}
\binom{k+1}{\ell}X^{\underline{\ell}}Y^{\underline{k+1-\ell}}\right] \\
&  =E\left[  \sum_{\ell=0}^{k+1}\left(  \binom{k}{\ell-1}+\binom{k}{\ell}\right)
X^{\underline{\ell}}Y^{\underline{k+1-\ell}}\right] \\
&  =\sum_{\ell=0}^{k+1}\binom{k}{\ell-1}E\left[  X^{\underline{\ell}}
Y^{\underline
{k+1-\ell}}\right]  +\sum_{\ell=0}^{k+1}\binom{k}{\ell}
E\left[  X^{\underline{\ell}%
}Y^{\underline{k+1-\ell}}\right] \\
&  =\sum_{\ell=0}^{k}\binom{k}{\ell}
E\left[  X^{\underline{\ell+1}}\right]  E\left[
Y^{\underline{k-\ell}}\right]  +\sum_{\ell=0}^{k}
\binom{k}{\ell}E\left[  X^{\underline
{\ell}}\right]  E\left[  Y^{\underline{k+1-\ell}}\right] \\
&  \leq\sum_{\ell=0}^{k}\binom{k}{\ell}\lambda 
E\left[  X^{\underline{\ell}}\right]
E\left[  Y^{\underline{k-\ell}}\right]  +\sum_{\ell=0}^{k}\binom{k}{\ell}
E\left[
X^{\underline{\ell}}\right]  \mu E\left[  Y^{\underline{k-\ell}}\right] \\
&  =(  \lambda+\mu)  E[  (  X+Y)  ^{\underline{k}%
}]  .
\end{align*}
The second property is again easily checked using Lemma~\ref{lem:scalmom}.%
{\hspace*{\fill}~\IEEEQED\par\endtrivlist\unskip}

\medskip

{\em Proof of Theorem~\ref{thm:ltnstrong}:}
In order to apply Proposition~\ref{prop:LlogL}
with $P^{*n}$ in place of $P$ and $\alpha=1/n$,
we need to check that 
$D(P^{*n}\|\Po(n\lambda))$ is finite.
Let $S_n$ denote the sum of $n$ i.i.d.\
random variables $X_i\sim P$, so that
$P^{*n}$ is the distribution of $S_n$.
Similarly, $\Po(n\lambda)$
is the sum of $n$ independent
$\Po(\lambda)$ variables. Therefore,
using the data-processing inequality 
\cite{csiszar-shields:FnT}
as in \cite{KonHarJoh03} implies that 
$D(P^{*n}\|\Po(n\lambda))\leq n D(P\|\Po(\lambda))$,
which is finite by assumption.

Proposition~\ref{prop:LlogL} gives,
$$D(T_{1/n}(P^{*n})\|\Po(\lambda))
\leq
	\frac{1}{2n^2(1-1/n)}
	+E[(S_n/n)\log(S_n/n)]
	-\lambda\log\lambda.
$$
By the law of large numbers, $S_n/n\to\lambda$
a.s., so
$(S_n/n)\log(S_n/n)\to\lambda\log\lambda$
a.s., as $n\to\infty$. Therefore,
to complete the proof it suffices 
to show that 
$(S_n/n)\log(S_n/n)$ converges to 
$\lambda\log\lambda$ also in $L^1$, 
or, equivalently, that the
sequence $\{T_n=(S_n/n)\log(S_n/n)\}$ is
uniformly integrable. We will actually
show that the nonnegative 
random variables $T_n$ are bounded above 
by a different uniformly integrable sequence.
Indeed, by the log-sum inequality,
\be
T_n
&=&
	\sum_{i=1}^n\frac{X_i}{n}\log
	\left(
	\frac{\sum_{i=1}^n\frac{X_i}{n}}
	{\sum_{i=1}^n\frac{1}{n}}
	\right)
	\nonumber\\
&\leq&
	\frac{1}{n}\sum_{i=1}^nX_i\log X_i.
	\label{eq:lln}
\ee
Arguing as in the beginning of the proof of
Proposition~\ref{prop:LlogL} shows that 
the mean $\mu=E[X_i\log X_i]$ is finite, 
so the law of large numbers implies that the averages
in (\ref{eq:lln}) converge to $\mu$
a.s.\ and in $L^1$. Hence, they
form a uniformly integrable sequence;
this implies that the $T_n$ are also
uniformly integrable, completing the proof.
{\hspace*{\fill}~\IEEEQED\par\endtrivlist\unskip}

\medskip

{\em Proof of Theorem~\ref{thm:ltnstrong2}:}
The proof is similar to that of Theorem~\ref{thm:ltnstrong},
so some details are omitted.
For each $n\geq 1$,
let $\lambda^{(n)}=\frac{1}{n}\sum_{i=1}^n\lambda_i$
and write $S_n=\sum_{i=1}^nX_i,$ 
where the random variables $X_i$ 
are independent, with each $X_i\sim P_i$.

First, to see that 
$D(P^{(n)}\|\Po(n\lambda^{(n)}))$ is finite,
applying the data-processing inequality 
\cite{csiszar-shields:FnT}
as in \cite{KonHarJoh03} gives,
$D(P^{(n)}\|\Po(n\lambda^{(n)}))\leq \sum_{i=1}^nD(P_i\|\Po(\lambda_i))$,
and it is easy to check that each of these terms
is finite because all $P_i$ have finite second moments.
As before, Proposition~\ref{prop:LlogL} gives,
\be
D(T_{1/n}(P^{(n)})\|\Po(\lambda^{(n)}))
\leq
	\frac{1}{2n^2(1-1/n)}
	+E[(S_n/n)\log(S_n/n)]
	-\lambda^{(n)}\log\lambda^{(n)}.
\label{eq:fromP}
\ee
Letting $Y_i=X_i-\lambda_i$ for each $i$,
the independent random variables $Y_i$ have 
zero mean and,
$$
\sum_{i=1}^\infty\frac{1}{i^2}E(Y_i^2)
=
\sum_{i=1}^\infty\frac{1}{i^2}\var(X_i^2)
\leq
\sum_{i=1}^\infty\frac{1}{i^2}E(X_i^2),
$$
which is finite by assumption~(b).
Then, 
by the general version of the law of large 
numbers on \cite[p.~239]{FellerII71}, 
$\frac{1}{n}\sum_{i=1}^nY_i\to 0$, 
a.s., and hence, by assumption~(a),
$S_n/n\to\lambda$
a.s., so that also,
$(S_n/n)\log(S_n/n)\to\lambda\log\lambda$
a.s., as $n\to\infty$. Moreover,
since $(x\log x)^{4/3}\leq x^2$ for
every integer $x\geq 1$, we have,
\ben
E\left\{\Big( \frac{S_n}{n}\log\frac{S_n}{n}\Big)^{4/3}\right\}
&\leq&
	E\left\{\Big( \frac{S_n}{n}\Big)^2\right\}\\
&=&
	\frac{1}{n^2}\sum_{i=1}^nE(X_i^2)
	+\frac{1}{n^2}\sum_{1\leq i\neq j\leq n} E(X_i)E(X_j)\\
&\leq&
	\sum_{i=1}^n
	\frac{1}{i^2}
	E(X_i^2)
	+(\lambda^{(n)})^2,
\een
which is uniformly bounded over $n$ by our
assumptions. Therefore, 
the sequence $\{(S_n/n)\log(S_n/n)\}$ 
is bounded in $L^p$ with $p=4/3>1$, which implies
that it is uniformly integrable,
therefore it converges to 
$\lambda\log\lambda$ also in $L^1$,
so that,
$D(T_{1/n}(P^{(n)})\|\Po(\lambda^{(n)}))\to 0$
as $n\to\infty$. 

Finally, recalling once more that 
the Poisson measures form
an exponential family, 
they satisfy a Pythagorean identity 
\cite{csiszar-shields:FnT},
so that 
\ben
D(T_{1/n}(P^{(n)})\|\Po(\lambda))
&=&
D(T_{1/n}(P^{(n)})\|\Po(\lambda^{(n)}))
+ D(\Po(\lambda^{(n)})\|\Po(\lambda)),
\een
where the first term was just shown to
go to zero as $n\to\infty$, and the
second term is actually equal to,
$$
\lambda^{(n)}
\log\frac{\lambda^{(n)}}{\lambda}
+\lambda - \lambda^{(n)},
$$
which also vanishes as $n\to\infty$
by assumption~(a).
{\hspace*{\fill}~\IEEEQED\par\endtrivlist\unskip}

\medskip

{\em Proof of Proposition~\ref{Charlierexpo}:}
Let $X_\alpha$ and $Z$ denote independent
random variables with distributions
$T_\alpha P$ and $\Po((1-\alpha)\lambda)$,
respectively. Then from the definitions,
and using Lemmas~\ref{lem:fallfact} and~\ref{lem:scalmom},
\ben
E[P_k^\lambda(X_{\alpha,\lambda})]
&=&
	\frac{1}{(\lambda^k k!)^{1/2}}
	\sum_{\ell=0}^k
	(-\lambda)^{k-\ell}
	\binom{k}{\ell}
	E\Big\{
	(X_\alpha+Z)^{\underline{\ell}}
	\Big\}\\
&=&
	\frac{1}{(\lambda^k k!)^{1/2}}
	\sum_{\ell=0}^k
	(-\lambda)^{k-\ell}
	\binom{k}{\ell}
	\Big\{
	\sum_{m=0}^\ell
	\binom{\ell}{m}
	E(X_\alpha^{\underline{m}})
	E(Z^{\underline{\ell-m}})
	\Big\}\\
&=&
	\frac{1}{(\lambda^k k!)^{1/2}}
	\sum_{\ell=0}^k
	(-\lambda)^{k-\ell}
	\binom{k}{\ell}
	\Big\{
	\sum_{m=0}^\ell
	\binom{\ell}{m}
	\alpha^m
	E(X^{\underline{m}})
	((1-\alpha)\lambda)^{\underline{\ell-m}}
	\Big\},
\een
where we have used the fact that the 
factorial moments of a
$\Po(t)$ random variable $Z_t$ satisfy,
$E[Z_t^{\underline{n}}]=t^n$. Simplifying
and interchanging the two sums,
\ben
E[P_k^\lambda(X_{\alpha,\lambda})]
&=&
	\frac{1}{(\lambda^k k!)^{1/2}}
	\sum_{m=0}^k
	\binom{k}{m}
	\alpha^m
	E(X^{\underline{m}})
	\sum_{\ell=m}^k
	\binom{k-m}{\ell-m}
	(-\lambda)^{k-\ell}
	((1-\alpha)\lambda)^{\underline{\ell-m}}\\
&=&
	\frac{1}{(\lambda^k k!)^{1/2}}
	\sum_{m=0}^k
	\binom{k}{m}
	\alpha^m
	E(X^{\underline{m}})
	(-\alpha\lambda)^{k-m}\\
&=&
	\alpha^kE[P_k^\lambda(X)],
\een
as claimed.
{\hspace*{\fill}~\IEEEQED\par\endtrivlist\unskip}

\medskip

{\em Proof of Proposition~\ref{prop:radon}:}
First we have to prove that $P/\Po(\lambda)\in L^2$.
Assume $P$ is Poisson bounded with ration $\mu$, say.
Using the bound in Lemma~\ref{lem:pbb},
\begin{align*}
\sum_{x=0}^{\infty}\Po(\lambda,x)\left(\frac{P(x)}{\Po(\lambda,x)}\right)^2  
& \leq
\sum_{x=0}^{\infty}\Po(\lambda,x)\left(\frac{\Po(\mu,x)\,e^\mu}
	{\Po(\lambda,x)}\right)^2\\
& =e^{\lambda}\sum_{x=0}^{\infty}\frac{(\mu^2/\lambda)^x}{x!}\\
& =e^{\lambda+\mu^{2}/\lambda},
\end{align*}
which is finite.

Now, recalling the general expansion (\ref{eq:PCexpand}),
it suffices to show that 
$\langle P/\Po(\lambda),P_k^\lambda\rangle=E[P_k^\lambda(X)]$.
Indeed, for $Z\sim\Po(\lambda)$,
$$
\Big \langle \frac{P}{\Po(\lambda) },P_{k}^{\lambda}\Big\rangle 
=
E\left( \frac{P(Z)}{\Po(\lambda,Z)} P_{k}^{\lambda}(Z)\right)
=
E\left[  P_{k}^{\lambda}\left(  X\right)  \right],
$$
as required.
{\hspace*{\fill}~\IEEEQED\par\endtrivlist\unskip}

\medskip

{\em Proof of Proposition~\ref{prop:altsum}:}
We need the following simple lemma; for a proof
see, e.g., \cite{gerber}.

\medskip

\begin{lemma}
If%
\[
F\left(  m,x\right)  =\sum_{\ell=0}^{m}\binom{x}{\ell}\left(  -1\right)^{\ell}%
\]
then%
\[
\left\{
\begin{array}
[c]{cc}%
F\left(  m,x\right)  \geq\delta_{x} & \text{for }m\text{ even,}\\
F\left(  m,x\right)  \leq\delta_{x} & \text{for }m\text{ odd.}%
\end{array}
\right. 
\]
\end{lemma}

\medskip

\noindent
Turning to the proof of Proposition~\ref{prop:altsum},
assume $X\sim P$ is Poisson bounded with ratio $\lambda$.
Then the series in the statement converges, since%
\[
\frac{1}{x!}\sum_{\ell=0}^{\infty}\left|  \left(  -1\right)^{\ell}\frac{E\left[
X^{\underline{x+\ell}}\right]  }{\ell!}\right|  \leq\,\frac{1}{x!}
\sum_{\ell=0}%
^{\infty}\frac{\lambda^{x+\ell}}{\ell!}=\Po(\lambda,x)
<\infty.
\]
For $m$ even we have,
\[
\delta_{z-x}\leq\sum_{\ell=0}^{m}\binom{z-x}{\ell}\left(  -1\right)  ^{\ell},%
\]
therefore,
\[
\binom{z}{x}\delta_{z-x}\leq\sum_{\ell=0}^{m}\binom{z}{x}\binom{z-x}{\ell}\left(
-1\right)  ^{\ell}=\frac{1}{x!}\sum_{\ell=0}^{m}\left(  -1\right)  ^{\ell}%
\frac{z^{\underline{x+\ell}}}{\ell!}\;.
\]
Multiplying by $P(z)$ and summing over $z$,
\begin{align*}
P(z)   
&  =\sum_{z=0}^{\infty}P(z)\binom{z}{x}\delta_{z-x}\\
&  \leq\sum_{z=0}^{\infty}P(z)  \frac{1}{x!}\sum_{\ell=0}%
^{m}\left(  -1\right)  ^{\ell}\frac{z^{\underline{x+\ell}}}{\ell!}\\
&  =\frac{1}{x!}\sum_{\ell=0}^{m}\left(  -1\right)  ^{\ell}\frac{E\left[
X^{\underline{x+\ell}}\right]  }{\ell!}\;.
\end{align*}
A similar argument holds for $m$ odd.
{\hspace*{\fill}~\IEEEQED\par\endtrivlist\unskip}

\medskip

{\em Proof of Theorem~\ref{thm:kthin}:}
Let $X_\alpha$ have distribution $T_\alpha P$. 
Using Lemma~\ref{lem:scalmom},
Proposition~\ref{prop:altsum},
and the fact that $X$ is ultra bounded,
the score function of $X_\alpha$ can be bounded as,
\be
\rho_{X_{\alpha}}(z)   
&=&
	\frac{(z+1)T_{\alpha}P(z+1)}{\alpha\lambda T_{\alpha}P(z)  }-1
	\nonumber\\
&\leq&
	\frac{(z+1)  
	E[  X_{\alpha}^{\underline{z+1}}]/
	(z+1)!
	}
	{\alpha\lambda
		\left( 
			 E[X_{\alpha}^{\underline{z}}]
	-E[X_{\alpha}^{\underline{z+1}}]%
	\right)/z!  }
	-1
	\nonumber\\
&=&
	\frac{\alpha^{z+1}E[X^{\underline{z+1}}]}
         {\alpha\lambda\left(  \alpha^{z}E[X^{\underline{z}}]  
	-\alpha^{z+1}E[X^{\underline{z+1}}]\right)  }-1
	\nonumber\\
&=& 
	\left[\frac{\lambda E[X^{\underline{z}}]}
	{E[X^{\underline{z+1}}]
	}-\lambda\alpha\right]^{-1}-1
	\nonumber\\
&\leq&
	\left[1-\lambda\alpha\right]^{-1}-1
	\nonumber\\
&=&
	\frac{\alpha\lambda}{1-\alpha\lambda}.
	\nonumber
\ee
Since the lower bound $\rho_{X_\alpha}(z)\geq-1$ is obvious,
it follows that,
\be
\rho_{X_\alpha}(z)^2\leq 1,\;\;\;\;\mbox{for all $\alpha>0$ small enough.}
\label{eq:scoreU}
\ee

We express $K(T_\alpha P)$ in three terms:
\be
K(T_\alpha P)=
\lambda\alpha\sum_{z=0}^{\kappa-2} T_\alpha P(z)\rho_{X_\alpha}(z)^2
+\lambda\alpha T_\alpha P(\kappa-1)\rho_{X_\alpha}(\kappa-1)^2
+\lambda\alpha\sum_{z=\kappa}^\infty T_\alpha P(z)\rho_{X_\alpha}(z)^2.
\label{eq:three-terms}
\ee
For the third term note that, applying Markov's inequality to 
the function $f(x)=x(x-1)\cdots (x-\kappa+1),$ which increases 
on the integers, we obtain,
\[
\Pr\{X_{\alpha}\geq\kappa\}\leq\frac{E[X_{\alpha}^{\underline{\kappa}}]}
{\kappa!}=\frac{\alpha^{\kappa}E[X^{\underline{\kappa}}]}{\kappa!}
\leq\frac{(\alpha\lambda)^{\kappa}%
}{\kappa!}.
\]
Therefore, using this and (\ref{eq:scoreU}),
for small enough $\alpha>0$ the third term 
in (\ref{eq:three-terms}) is 
bounded above by,
$$\alpha\lambda\frac{(\alpha\lambda)^{\kappa}}{\kappa!}\to 0,$$
which, divided by $\alpha^\kappa$, tends to zero 
as $\alpha\to 0$.

For the other two terms we use the full expansion of
Proposition~\ref{prop:altsum}, together
with Lemma~\ref{lem:scalmom}, to obtain a more accurate
expression for the score function,
\ben
\rho_{X_\alpha}(z)
&=&
	\frac{(z+1)T_\alpha P(z+1)-\alpha\lambda T_\alpha P(z)}
	{\alpha\lambda T_\alpha P(z)}
	\\
&=&
	\frac{\frac{1}{z!}\sum
	_{\ell=0}^{\infty}\left(  -1\right)  ^{\ell}\frac{E[ 
	X_{\alpha}^{\underline{z+1+\ell}}]  }{\ell!}
	-\alpha\lambda\frac{1}{z!}%
	\sum_{\ell=0}^{\infty}\left(  -1\right)  ^{\ell}
	\frac{E[X_{\alpha}^{\underline{z+\ell}}]}{\ell!}}
	{\alpha\lambda\frac{1}{z!}%
	\sum_{\ell=0}^{\infty}\left(  -1\right)  ^{\ell}
	\frac{E[X_{\alpha}^{\underline{z+\ell}}]  }{\ell!}}\\
&=&
	\frac{\sum_{\ell=0}^{\infty}\left(  -1\right)  ^{\ell}\alpha^{\ell}
	\left(  E\left[  X^{\underline{z+1+\ell}}\right]  -\lambda E\left[
	X^{\underline{z+\ell}}\right]  \right)/\ell!}
	{\lambda\sum_{\ell=0}^{\infty}\left(
	-1\right)^{\ell}\alpha^{\ell}
	E[  X^{\underline{z+\ell}}]/\ell!}.
\een
Since, by assumption,
$E\left[  X^{\underline{z+1+\ell}}\right]  
-\lambda E\left[  X^{\underline{z+\ell}
}\right]  =0$ for $z+\ell<\kappa-1$,
the first terms in the series in the numerator above
vanish.  Therefore,
$$\rho_{X_\alpha}(z)
=
\alpha^{\kappa-z-1}
\frac{\sum_{\ell=0}^{\infty}\left(  -1\right)
^{\ell+\kappa-z-1}\alpha^{\ell}
\left(
E\left[  X^{\underline{\ell+z}}\right]  
-\lambda E\left[  X^{\underline{\ell+z-1}%
}\right]  \right)/(\ell+\kappa-z-1)!}
{\lambda\sum_{\ell=0}^{\infty}\left(  -1\right)  ^{\ell}%
\alpha^{\ell}E[  X^{\underline{z+\ell}}]/\ell!}.
$$
For $z\leq\kappa-2$, the numerator and denominator above
are both bounded functions of $\alpha$, and the denominator
is bounded away from zero (because of the term corresponding
to $\ell=0$). Therefore, for each $0\leq z\leq\kappa-2$,
the score function $\rho_{X_\alpha}(z)$ is of order
$\alpha^{\kappa-z-1}$. For the first term in 
(\ref{eq:three-terms}) we thus have,
\ben
\lambda\alpha\sum_{z=0}^{\kappa-2} T_\alpha P(z)\rho_{X_\alpha}(z)^2
=
	\alpha \sum_{z=0}^{\kappa-2} 
	O(\alpha^z)
	O(\alpha^{2\kappa-2z-2})
=
	O(\alpha^{\kappa+1}),
\een
which, again, when divided by $\alpha^\kappa$, tends to zero 
as $\alpha\to0.$

Thus only the second term in (\ref{eq:three-terms})
contributes. For this term, we similarly obtain,
\be
\lim_{\alpha\rightarrow0}\rho_{X_{\alpha}}(\kappa-1)   
&=&
	\lim_{\alpha\rightarrow0}
	\frac{\sum_{j=0}^{\infty}(-1)^j\alpha^j
	\left(  E\left[X^{\underline{j+\kappa}}\right]  
		-\lambda E\left[ X^{\underline{j+\kappa-1}}\right]  \right)/j!}
	{\lambda\sum_{j=0}^\infty(-1)^j
	\alpha^jE\left[X^{\underline{j+\kappa-1}}\right]/j!  }
	\nonumber\\
&=&
	\frac{E\left[  X^{\underline{\kappa}}\right]  -\lambda E\left[
	X^{\underline{\kappa-1}}\right]  }
	{\lambda E\left[  X^{\underline{\kappa-1}%
	}\right]  }
	\nonumber\\
&=&
	\frac{E\left[  X^{\underline{\kappa}}\right]  
	-\lambda^{\kappa}%
	}{\lambda^{\kappa}},
	\label{eq:exchange1}
\ee
and,
\be
\lim_{\alpha\rightarrow0}\frac{\alpha\lambda T_{\alpha}P(\kappa-1)}
{\alpha^{\kappa}}
&=&
	\lim_{\alpha\rightarrow0}\frac
	{\alpha\lambda
	\sum_{j=0}^{\infty}\left(
	-1\right)  ^{j}E\left[
	(T_{\alpha}X)^{\underline{\kappa-1+j}}
	\right]  /j!}{(\kappa-1)!\,\alpha^{\kappa}}
	\nonumber\\
&=&
	\lim_{\alpha\rightarrow0}
	\frac{\lambda
	\sum_{j=0}^{\infty}\left(
	-1\right)  ^{j}
	\alpha^j
	E\left[
	X^{\underline{\kappa-1+j}}
	\right] / j!}{(\kappa-1)!}
	\nonumber\\
&=&
	\frac{\lambda^{\kappa}}{\left(
	\kappa-1\right)  !}.
	\label{eq:exchange2}
\ee
Finally, combining the above limits 
with (\ref{eq:three-terms}) yields,
\begin{align*}
\lim_{\alpha\rightarrow0}\frac{K\left(  T_{\alpha}X\right)  }{\alpha^{\kappa}}
&  =\frac{\lambda^{\kappa}}{\left(  \kappa-1\right)  !}\left(  \frac{E\left[
X^{\underline{\kappa}}\right]  -\lambda^{\kappa}}{\lambda^{\kappa}}\right)
^{2}\\
&  =\kappa\left(  \frac{E\left[  X^{\underline{\kappa}}\right]  -\lambda
^{\kappa}}{\lambda^{\kappa/2}\left(  \kappa!\right)  ^{1/2}}\right)  ^{2}\\
&  =\kappa E\left[  P_{\kappa}\left(  X\right)  \right]  ^{2},
\end{align*}
as claimed.
{\hspace*{\fill}~\IEEEQED\par\endtrivlist\unskip}

\medskip

{\em Proof of Lemma~\ref{lem:poincarish}:}
The key is to observe that for $Y$ ULC, since $P(y+1) 
(y+1)/P(y)$ is decreasing
in $y$, and $y$ is increasing in $y$, there exists an integer $y_0$ such
that $P(y+1)(y+1) \leq y_0 P(y)$ for $y \geq y_0$ and
$P(y+1) (y+1) \geq y_0 P(y)$ for $y < y_0$. Hence:
\begin{eqnarray*}
\sum_{y=z+1}^{\infty} P(y)(y-y_0)
&=&
	P(z+1) (z+1) + \sum_{y=z+1}^\infty 
	(P(y+1) (y+1) - y_0 P(y))\\
&\leq& 
	(z+1) P(z+1),\;\;\mbox{for}\;z \geq y_0;\\
\sum_{y=0}^{z} P(y)(y_0-y) 
&=&
	P(z+1) (z+1) - \sum_{y=0}^z
	( P(y+1) (y+1) - y_0 P(y))\\
&\leq&
	(z+1) P(z+1),
	\;\;\mbox{for}\; z \leq y_0-1.
\end{eqnarray*}
Further, by Cauchy-Schwarz, for $y \geq y_0$,
\begin{equation} \label{eq:cs} \left( g(y) - g(y_0) \right)^2 = 
\left( \sum_{z=y_0}^{y-1} \Delta g(z) \right)^2
\leq (y- y_0) \left( \sum_{z=y_0}^{y-1} \Delta g(z)^2 \right),
\end{equation}
while for $y \leq y_0-1$,
\begin{equation} \label{eq:cs2} \left( g(y) - g(y_0) \right)^2 = 
\left( \sum_{z=y}^{y_0-1} \Delta g(z) \right)^2
\leq (y_0- y) \left( \sum_{z=y_0}^{y-1} \Delta g(z)^2 \right).
\end{equation}
This means that (with the reversal of order of summation justified by Fubini,
since all the terms have the same sign),
\begin{eqnarray}
\lefteqn{\sum_{y=0}^{\infty} P(y) (g(y) - \mu)^2 } \nonumber \\
 & \leq & \sum_{y=0}^{\infty} P(y) \left( g(y) - g(y_0) \right)^2 \nonumber \\
& = & \sum_{y=0}^{y_0-1} P(y) \left( g(y) - g(y_0) \right)^2 +
\sum_{y=y_0}^{\infty} P(y) \left( g(y) - g(y_0) \right)^2 \nonumber \\
& \leq & \sum_{y=0}^{y_0-1} P(y)
(y_0- y) \left( \sum_{z=y}^{y_0-1} \Delta g(z)^2 \right) 
+ \sum_{y=y_0}^{\infty} P(y)
(y- y_0) \left( \sum_{z=y_0}^{y-1} \Delta g(z)^2 \right) 
\label{eq:step1} \\
& \leq & \sum_{z=0}^{y_0-1} \Delta g(z)^2 \left( \sum_{y=0}^{z} 
P(y) (y_0-y) \right) 
+ \sum_{z=y_0}^{\infty} \Delta g(z)^2 
\left( \sum_{y=z+1}^{\infty} P(y)(y-y_0) \right) 
\nonumber \\
& \leq & \sum_{z=0}^{\infty} (\Delta g)(z)^2 P(z+1) (z+1),  \label{eq:step2} 
 \end{eqnarray}
and the result holds. Note that the inequality 
in (\ref{eq:step1}) follows by (\ref{eq:cs}) and (\ref{eq:cs2}), 
and the inequality in (\ref{eq:step2}) by the
discussion above.
{\hspace*{\fill}~\IEEEQED\par\endtrivlist\unskip}

\bibliographystyle{plain}

\end{document}